\documentclass[useAMS,usenatbib,usegraphicx,epsfig]{aa}
\usepackage{txfonts,graphicx,epsfig}
\usepackage[figuresright]{rotating}
\def\eps@scaling{.95}
\def\epsscale#1{\gdef\eps@scaling{#1}}
\def\plotone#1{\centering \leavevmodes
\epsfxsize=\eps@scaling\columnwidth \epsfbox{#1}}

\newcommand{\nhi}{$N_{\rm \ion{H}{i}}$}
\newcommand{\nh}{$n_{\rm H}$}
\newcommand{\nos}{$N_{\rm \ion{O}{vi}}$}
\newcommand{\nciv}{$N_{\rm \ion{C}{iv}}$}
\newcommand{\nee}{\ion{Ne}{viii}}
\newcommand{\os}{\ion{O}{vi}}
\newcommand{\cf}{\ion{C}{iv}}
\newcommand{\ct}{\ion{C}{iii}}
\newcommand{\ctwo}{\ion{C}{ii}}
\newcommand{\ho}{\ion{H}{i}}
\newcommand{\sif}{\ion{Si}{iv}}
\newcommand{\sit}{\ion{Si}{iii}}
\newcommand{\nf}{\ion{N}{v}}
\newcommand{\nth}{\ion{N}{iii}}
\newcommand{\sus}{\ion{S}{vi}}
\newcommand{\icm}{cm$^{-2}$}
\newcommand{\hkpc}{$h_{70}^{-1}$ kpc}
\newcommand{\he}{HE1104$-$1805}
\newcommand{\rx}{RXJ0911.4$+$0551}

\newcommand{\kms}{km~s$^{-1}$}

\newcommand{\lya}{Ly$\alpha$\ }
\newcommand{\lyb}{Ly$\beta$\ }
\newcommand{\lyc}{Ly$\gamma$\ }

\newcommand{\lye}{Ly$\epsilon$\ }

\begin{document}

   \title{Clues to the nature of  high-redshift \ion{O}{vi} absorption
     systems \\ from their (lack of) small-scale structure\thanks{Based on
     observations 
     made at the  
     European Southern Observatory (ESO), under programs 67.A-0278(A) and
     70.A-0439(A), with the UVES spectrograph at the VLT, Paranal, Chile.}  }

\titlerunning{\os\ along double lines of sight}
\authorrunning{S. Lopez et al.}

   \author{S. Lopez,
          \inst{1}
          S. Ellison,\inst{2}
          S. D'Odorico,\inst{3}
          \and
          T.-S. Kim\inst{4,}\inst{5}
          }

   \offprints{S. Lopez}

   \institute{Departamento de Astronom\'ia, Universidad de Chile, Casilla
     36-D, Santiago, Chile \email{slopez@das.uchile.cl}    
   \and 
University of Victoria, Dept. Physics \& Astronomy, Elliott Building, 3800
Finnerty Rd, Victoria, V8P 1A1, British Columbia, Canada \email{sarae@uvic.ca}
\and
         European Southern Observatory, Karl-Schwarzschild-Str. 2, 85748
         Garching-bei-M\"unchen, Germany \email{sdodoric@eso.org} 
         \and
         Institute of Astronomy, Madingley Road, Cambridge CB3 0HA,
         UK  
         \and
Present address: Astrophysikalisches Institut Potsdam, 
An der Sternwarte 16, 
D-14482, Potsdam, Germany
\email{tkim@aip.de}
}

   \date{\today}

   \abstract{ We present results of the first survey of high-redshift
   ($<z>\sim 2.3$ ) \os\ absorption systems along parallel lines of sight
   toward two lensed QSOs. { After a careful and well-defined search,} we
   find ten intervening \os\ systems -- identified by the presence of { the
   $\lambda\lambda 1031,1037$ doublet lines}, \ho, and in most cases \cf,
   \sif, and \ct\ -- and eight candidate systems for which we do not detect
   \ho\ nor other metals. { We assess the veracity of these systems by
   applying a classification scheme.} { Within the errors,} all \os\
   systems appear at the same redshift and have similar line strengths in
   front of both QSO images, whereas in most cases \cf\ or \sif\ show more
   differences across the lines of sight, either in radial velocity or line
   strength.  We conclude that (1) the coherence length of \os\ must be much
   larger than $\approx 1$ \hkpc, and (2) an important fraction of the \cf\
   absorbers may not reside in the same volume as \os.  { Given the
   inhomogeneous character of the data --different S/N ratios and degrees of
   blending--, we pay special attention to the observational errors and their
   impact on the above conclusions.}  Since Doppler parameters are consistent
   with photoionization, we propose a model in which \cf\ occurs in two
   different photoionized phases, one large, with characteristic sizes of a
   few hundred kpc and bearing \os, and another one a factor of ten smaller
   and containing \ct. This model is able to explain the various transverse
   differences observed in column density and kinematics. We apply the model
   successfully to 2 kinds of absorbers, with low and high metallicity. In the
   low-metallicity regime, [C/H]$\sim -2$, we find that [C/O] $\approx -0.7$
   is required to explain the observations, which hints at late ($z\la 6$)
   rather than early metal enrichment. In the high-metallicity regime, the
   observed dissociation between \os\ and \cf\ gas might be produced by
   galactic outflows. Altogether, the relative abundances, inhomogeneous \cf\
   and featureless \os\ are consistent with gas that has been processed
   recently before the absorption occurred (thus close to star-forming
   regions).  Finally, we discuss briefly three associated systems ($z_{abs}
   \sim z_{em}$) that also show \os.

\keywords{cosmology: observations -- intergalactic medium -- 
galaxies: halos -- quasars: absorption lines -- quasars: individual:
\he, \rx }
               }
   \maketitle
%

\section{Introduction}

Although once supposed to contain primordial material unenriched by the
products of star formation, the \lya\ forest is now well known to frequently
exhibit associated metal lines.  The most commonly detected metal feature at
high redshift is the \cf\ doublet (e.g. Cowie \& Songaila 1998; Ellison et
al. 2000; Schaye et al. 2003; Aracil et al. 2004) due to the combination of
its elemental abundance, convenient rest wavelength and ionization potential.
Early models which used simple recipes for ionization corrections and assumed
uniform metallicities determined that the carbon abundance was in the range of
$10^{-2}$ to $10^{-3}$ of the solar value (e.g Haehnelt et al. 1996; Dav\'e et
al 1998).  More sophisticated modeling has confirmed this result, and shown
that the metallicity is most dependent on the gas overdensity, but largely
insensitive to redshift in the range $2 < z < 4$ (Schaye et al. 2003).
Although the low metallicities and relatively low \nhi\ column densities of
the associated \lya\ may hint toward an inter-galactic origin for \cf\
absorbers, recent work by Adelberger et al. (2003, 2005) has demonstrated a
strong connection between the high column density \cf\ systems and Lyman break
galaxies at $z \sim 3$.  Indeed, Adelberger et al. (2005) infer that
star-forming galaxies in their sample exhibit \cf\ absorption out to 80 kpc,
sufficient to account for the bulk of observed absorbers (but note that these
observations do not rule out an inter-galactic origin of the {\it weak} \cf\
absorbers detected with pixel-statistics techniques; Pieri, Schaye \& Aguirre
2006).


\begin{table*}
\begin{minipage}[t]{18cm}
\caption{VLT UVES Observations of \he\ and \rx\ }
\label{table_obs}      
\centering          
\renewcommand{\footnoterule}{}
         \begin{tabular}{lcccrcccc}
\hline\hline    
QSO&{\rm UVES Setup}&{\rm Wavelength}&{\rm Exp. Time}&{\rm Observing Date}&{\rm
  Seeing}& FWHM\footnote{Resolution of final spectrum.} &
S/N(A)\footnote{Median signal-to-noise ratio per pixel blue and redwards of
  \lya\ emission.} & S/N(B)$^b$\\
&              &[nm]           &  [s]        &  & [arcsec]&\kms\ &&       \\
            \hline
\he&Dic- (390+580)& 328-445,476-684 &7\,600&Jun01&0.7--0.9&6.7&30,90&15,40\\ 
   &Dic- (437+860)  &376-500,667-1040&40\,020&Apr01--Jan02&0.5--1.0\\
   &Dic- (346+564)  &305-390,460,660&11\,400&Apr01--Jun01&0.5--0.8\\
\rx&Dic- (390+564)&328-445,460-560,567-660&43\,200&Dec02--Feb03&0.7--1.1&7.0&40,60&12,20\\ 
\hline                  
\end{tabular}
\end{minipage}
\end{table*}

With large
datasets of high S/N spectra, extensive catalogs of other metal
species such as \sif\ (e.g., Aguirre et al. 2004) and \os\ (e.g., Schaye
et al. 2000; Carswell, Schaye \& Kim 2002; Simcoe, Sargent \&
Rauch 2002, 2004; Levshakov et al. 2003) are now becoming available.  
Species which represent a range of nucleosynthetic origins and ionization
states allow us to probe a range of environments and physical states.
For example, \os\ is a particularly advantageous tracer of metals 
in the low density \lya\ forest  because it has the highest ionization 
fraction for \nhi $\la 10^{15}$ \icm\ (e.g. Schaye et al. 2000; Simcoe 
et al. 2004), and because oxygen is a cosmically
abundant element.  These \os\ studies have inferred enrichment of
$10^{-2}$--$10^{-3} Z_{\odot}$ in IGM structures close to the mean
density of the universe at $z \sim 2$.  However, such widespread
metal pollution is not necessarily indicative of in-situ Population
III star enrichment.  Indeed, a number of authors (e.g. Aguirre et al.
2001; Scannapieco et al 2002; Simcoe et al. 2004) have modeled the 
contributions from Pop III stars versus galactic winds and concluded 
that the latter is the dominant enrichment source.  The connection
between \os\ and galaxies has been strengthened further by correlations
between color-selected galaxies (many with spectroscopic redshifts) 
and absorbers by Simcoe et al. (2006).  Ionization modeling of these
\os\ absorbers, many of which exhibit a suite of other metal
species such as \nf, \ct\ and \sit, led Simcoe et al. (2006) to conclude
that the high metallicities, large coherence scales and small cloud
sizes were consistent with material being expelled from galaxies, e.g.
by supernova-driven winds.  However, others have
argued that early populations of massive stars contribute a significant
fraction of the metals, up to 60\% according to Qian \& Wasserburg (2005).

Further clues to the origin and environment of \os\ absorbers can be achieved
through ionization modeling which can constrain densities and heating
mechanisms.  So far, many such studies have been executed at \textit{low
redshift}. For example, modeling the population of $z \lesssim 0.3$ \os\
absorbers discovered by HST and FUSE have revealed predominantly collisionally
ionized gas at $T \sim 10^5$ K (e.g Tripp \& Savage 2000; Tripp, Savage \&
Jenkins 2000; Danforth \& Shull 2005; Stocke et al. 2006), although
multi-phase models are often required (Danforth et al. 2005).  The majority of
low redshift \os\ absorbers have therefore been associated with inter-galactic
gas that is shock heated as it forms large scale structures, the so-called
warm-hot intergalactic medium (WHIM).  \os\ has also now been detected in a
large number of Galactic sightlines, both in the disk (e.g. Bowen et al. 2005)
and in halo clouds (e.g.  Savage et al. 2003), and in high-velocity clouds
(Fox, Savage \& Wakker 2006).  The Galactic \os\ absorbers
have been associated with violent dynamical and non-equilibrium processes
(Indebetouw \& Shull 2004; Bowen et al. 2005).  In contrast, high redshift
\os\ absorbers have so far almost universally been found to be photo-ionized
(Bergeron et al. 2002; Carswell et al. 2002; Levshakov et al 2003, but see
Reimers et al. 2001, and Simcoe, Sargent \& Rauch 2002).

The Adelberger et al. (2005) sizes of LBG metal enrichment are
consistent with limits obtained from binary or gravitationally lensed
QSOs which can be used to determine the transverse size of absorbers.
A number of studies (e.g., Smette et al.\ 1995; Petitjean et al. 1998;
Lopez, Hagen \& Reimers 2000; Rauch, Sargent \& Barlow 2001) have seen
correlations on scales up to tens of kpc, but with fractional column
density variation increasing up to 50\% for 100 kpc separations
(e.g. Fig. 11 in Ellison et al. 2004).

In this paper we combine the utility of gravitational lensing with
ionization modeling to further investigate the physical state and
coherence scale of \os\ absorbers.  This is the first time that a 
direct measure of \os\ sizes from gravitational lensing has been 
achieved.  We complement our coherence scale limits derived from
lensing with models of one and two phase ionized media.  The combination
of constraints from these two directions allows us to derive
approximate sizes (under certain assumptions of the multi-phase
structure) and relevant abundances of C, N and O.

The targets that we use for this work are \he\ ($z_{em}=2.32$) and \rx\
($z_{em}=2.80$).  These objects have been studied extensively in the lensing
literature and, in the case of \he, studies on intervening absorption line
systems also exist (of which the most pertinent to our study are: Smette et
al. 1995; Lopez et al. 1999; Rauch, Sargent \& Barlow 2001). For \rx, optical
low-resolution spectra are reported in the discovery paper by Bade et
al. 1997, and medium-resolution spectra more recently by Rauch et al. (2005);
here we present the first high-resolution spectra.  Both lens systems have
transverse separations of $\approx 3\arcsec$. At the redshifts of the
intervening systems studied here, the line-of-sight (LOS) separation ranges
between $S=0.2$ and $1.4$ kpc, for a cosmology with $H_0=70$ \kms~Mpc$^{-1}$,
and $(\Omega_M,\Omega_\Lambda)=(0.3,0.7)$.

 The paper is organized as follows: after describing the observations in
Sect.~\ref{sect_obs}, in Sect.~\ref{sect_abs} we describe the selection method
and the individual \os\ absorption systems that were found, along with an
assessment of sample completeness.  In Sect.~\ref{comparisons} we analyze
observed properties of those systems across the lines of sight. In
Sect.~\ref{sect_photo} we describe results and uncertainties from
photoionization models. Finally, in Sect.~\ref{sect_metals} we compare
observations and models to draw inferences on the origin of the high-redshift
\os\ gas.  The 
conclusions are outlined in Sect.~\ref{conclusions} and in the Appendix we
describe the associated \os\ systems.


\section{Observations and data reduction}
\label{sect_obs}

We observed both QSO systems with the VLT UV-Visual Echelle Spectrograph
(UVES) in several runs from June 2001 through February 2003. Spectra of each
QSO system were acquired simultaneously by aligning the slit with the 2 lensed
images (``A'' and ``B''). This configuration is well defined for the double
QSO \he, but \rx\ is a quadruple system and what we shall call its ``A''
spectrum is actually the sum of three images (A1, A2, and A3 in Burud et
al. 1998) which are not resolved in normal seeing conditions.  Extraction of
each 2-dimensional spectra was carried out with our own pipeline, which is
based on simultaneous fitting of the spatial profiles (see Lopez et al. 2005
for details).  For both QSO systems images A and B are separated by 3--4
seeing units, so the the fitting procedure ran smoothly. Table~\ref{table_obs}
lists the observational and spectrum-quality parameters.

Before coadding orders, the spectra were rebinned to a common
vacuum-heliocentric scale with pixel size of $0.043$ \AA.  The final spectral
resolution in the coadded spectra was FWHM $6.7$ and $7.0$ \kms, for \he\ and
\rx, respectively.  Finally, the coadded orders were normalized by dividing
out the response function and QSO continuum simultaneously. The continuum was
estimated first in the echelle orders of the A spectra by fitting cubic
splines through featureless spectral regions.  The continuum in QSO B (the
faintest QSO component) was a scaled version of that of QSO A.  Minor
adjustments to the B continuum had to be made to account for differences in
the flux ratios on top of emission lines. In \rx, we have fitted out the
mini-broad absorption system at $z=2.57$ reported in Bade et al. (1997).

   \begin{figure}
   \centering
   \includegraphics[width=5.2cm,angle=-90,clip]{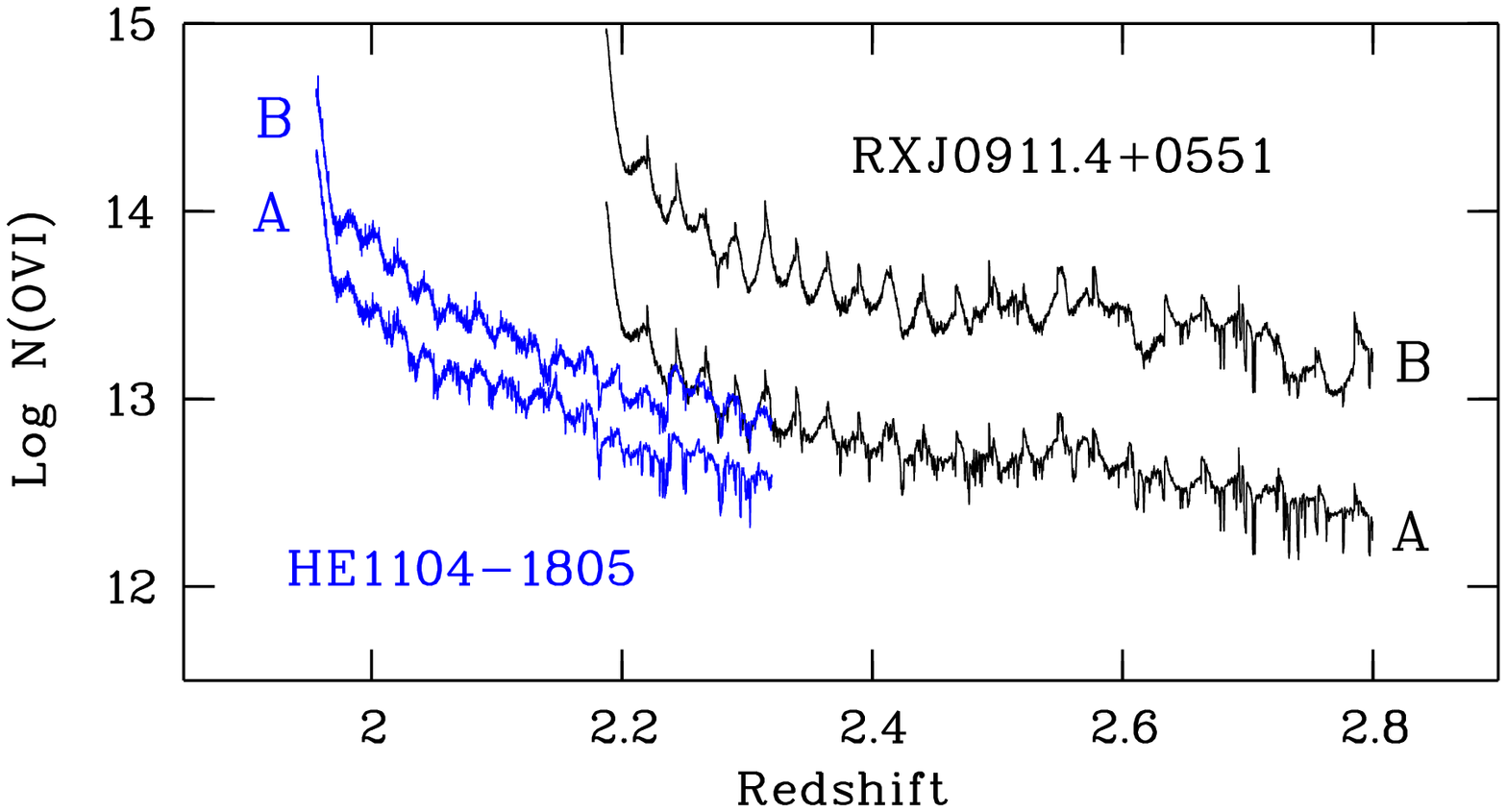}
   \caption{Three-$\sigma$ detection limits in terms of  \os\ column density
     for  \he\ and \rx.
     }
   \label{fig_detection} 
    \end{figure}

\begin{table*}
\begin{minipage}[t]{18cm}
\caption{Results of search for \os\ systems toward \he\ and \rx}\label{table_OVI}      
\begin{tabular}{ccccccc}
\hline\hline   

 QSO  &\multicolumn{2}{c}{LOS A}&& \multicolumn{2}{c}{LOS B}& Class\\
       \cline{2-3} \cline{5-6}

      & $z_{\rm A}$ & Method$^a$ && $z_{\rm B}$ & Method$^a$& \\ 
\hline
  \he\  & 2.314201  & 1,2 & &  2.314201& 1,2  & Associated\\
        & 2.299304  & 1,2 & &  2.299304& 1,2  & Associated\\
        & 2.200800  & 1,2 & &  2.200822& 1,2  & II\\
        & 2.158112  & 1,2 & &  2.158139& 1,2  & II\\
        & 2.086682  & 1,2 & &          &      & II\\
        &           &     & &  2.052440&   2  & II \\
        & 2.036100  & 3   & &          &      & III\\
        & 2.016083  & 3   & &          &      & III\\
        & 2.010473  & 2   & &  2.010452&   2  & II\\
  \rx\  & 2.776399  & 2   & &  2.776399& 2    & Associated \\
        &           &     & &  2.716287& 3    & III\\
        & 2.632495  & 1,2 & &  2.632495& 1,2  & I\\
        & 2.626749  &  1  & &          &      & II\\
        & 2.623886  &  1  & &          &      & II\\
        &           &     & &  2.602501& 3    & III\\
        & 2.577051  & 3   & &  2.573192& 3    & III\\
        & 2.517045  & 1,2 & &  2.516961& 2    & I\\
        & 2.437271  & 1,2 & &  2.437265& 2    & I\\
        & 2.431602  & 3   & &          &      & III\\
        & 2.418109  & 1,2 & &  2.418124& 1    & III\\
        & 2.238486  & 3   & &          &      & III\\	   
	 
\hline

\end{tabular}
\end{minipage}
\begin{list}{}{}
\item[$^a$] Method 1: \ho\ as a signpost; method 2: \cf\ as a
  signpost; method 3: blind \os\ search.
\end{list}
\end{table*}


   \begin{figure*}
   \centering
   \includegraphics[width=12cm,clip]{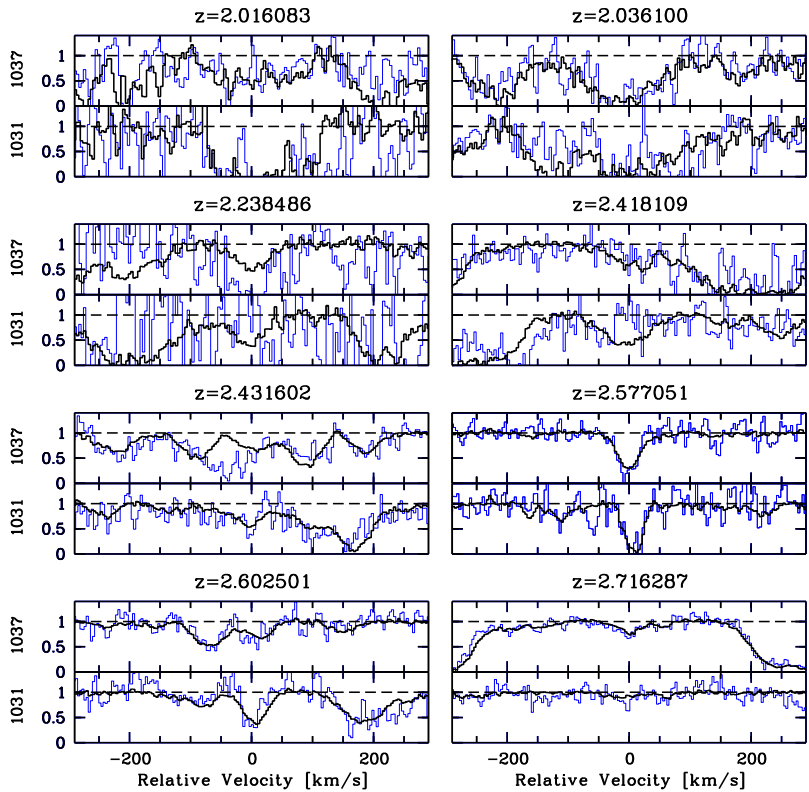}
   \caption{Class III \os\ systems toward \he\ and \rx\
   (see Table~\ref{table_OVI}). The thick black line corresponds to the A
   spectra; the thin blue one to the B spectra.  }   
   \label{fig_ClassIII}  
    \end{figure*}

   \begin{figure*}
   \centering
   \includegraphics[width=12cm,clip]{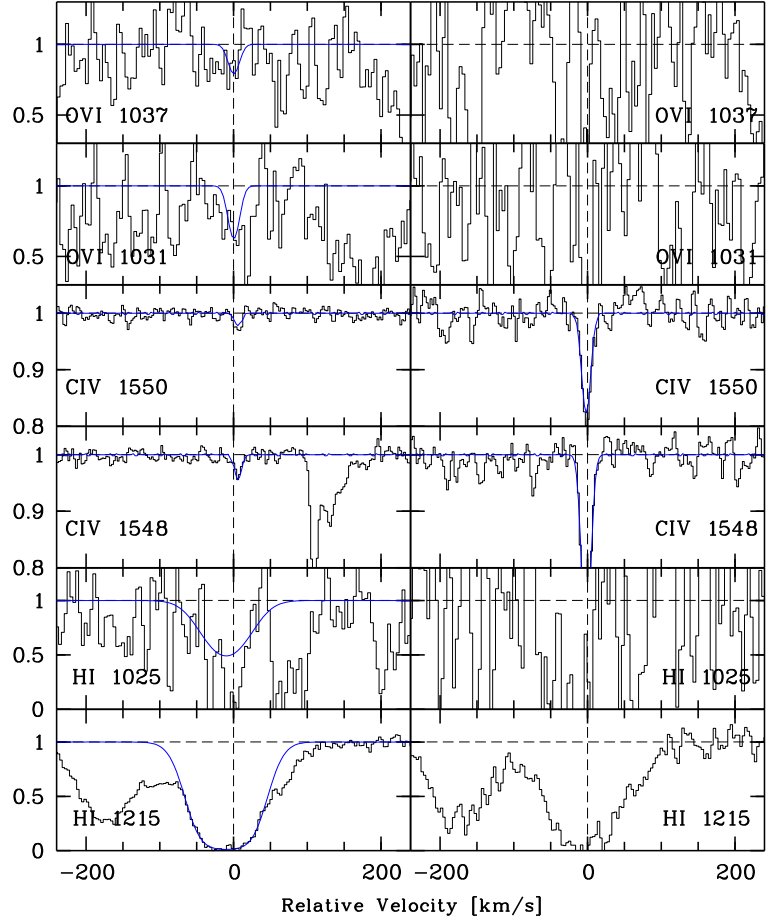}
   \caption{Class II system in \he. Lefthand (righthand) panels show
     transitions observed along sightline A (B). The smoothed lines show Voigt
     profiles with line parameters listed in Table~\ref{table_AB}. In both
     panels the vertical dashed line indicates $z=2.010473$, which is the
     redshift of \os\ in the A spectrum.}
   \label{fig_spec0} 
    \end{figure*}
   \begin{figure*}
   \centering
   \includegraphics[width=12cm,clip]{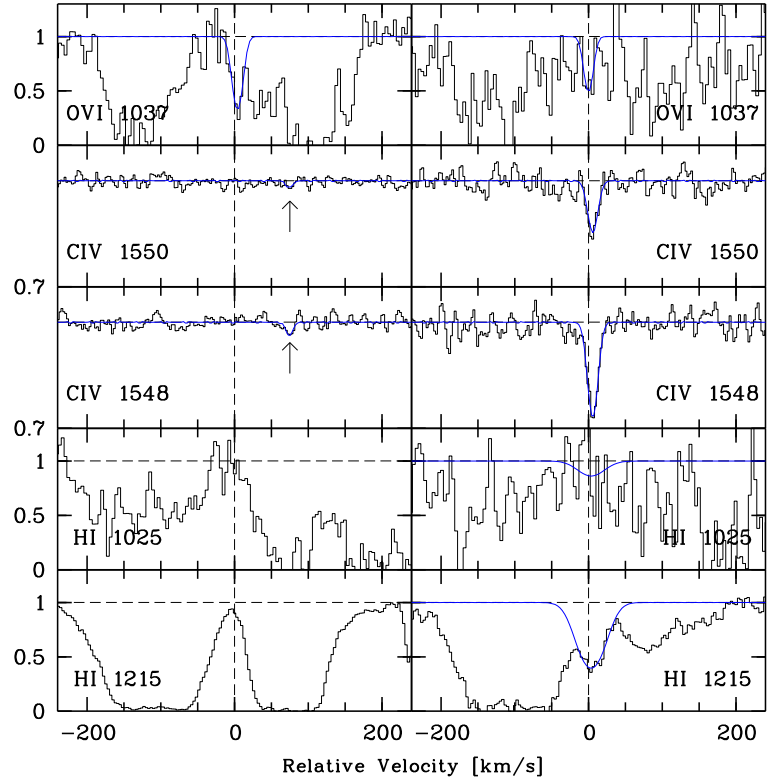}
   \caption{Class II system in \he. Panels and symbols as for
   Fig.~\ref{fig_spec0}, but for 
   $z=2.052440$, which is the
     redshift of \os\ in the B spectrum. Arrows mark a possible \cf\ system in
   the A spectrum.}  
   \label{fig_spec1a} 
    \end{figure*}

   \begin{figure*}
   \centering
   \includegraphics[width=12cm,clip]{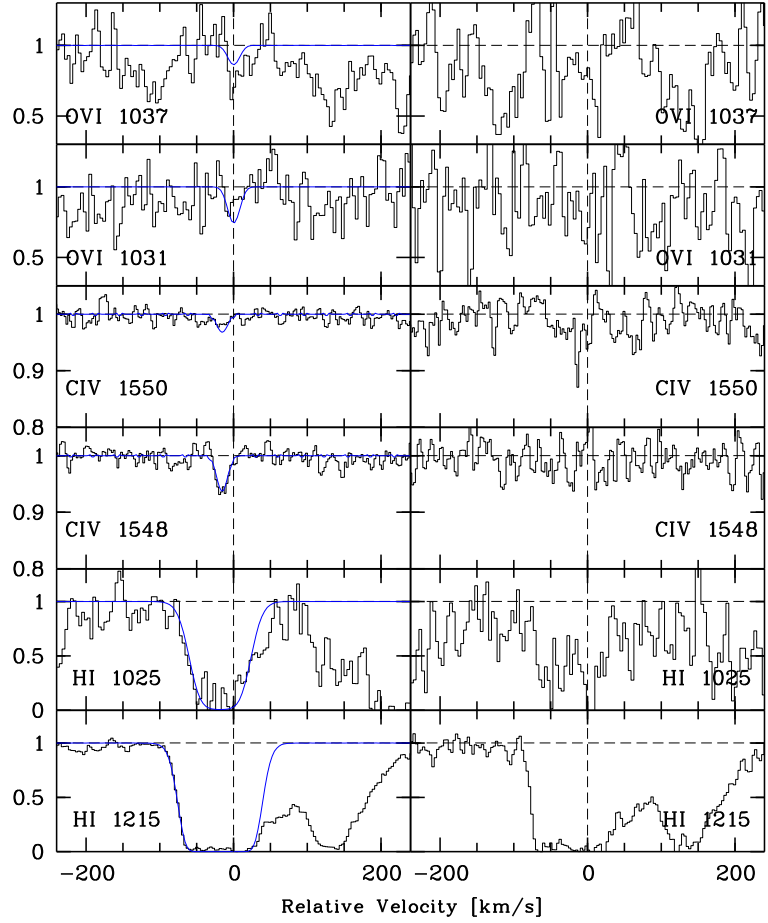}
   \caption{Class II system in \he. Panels and symbols as for
   Fig.~\ref{fig_spec0}, but for 
   $z=2.086682$.}  
   \label{fig_spec1} 
    \end{figure*}

   \begin{figure*}
   \centering
   \includegraphics[width=12cm,clip]{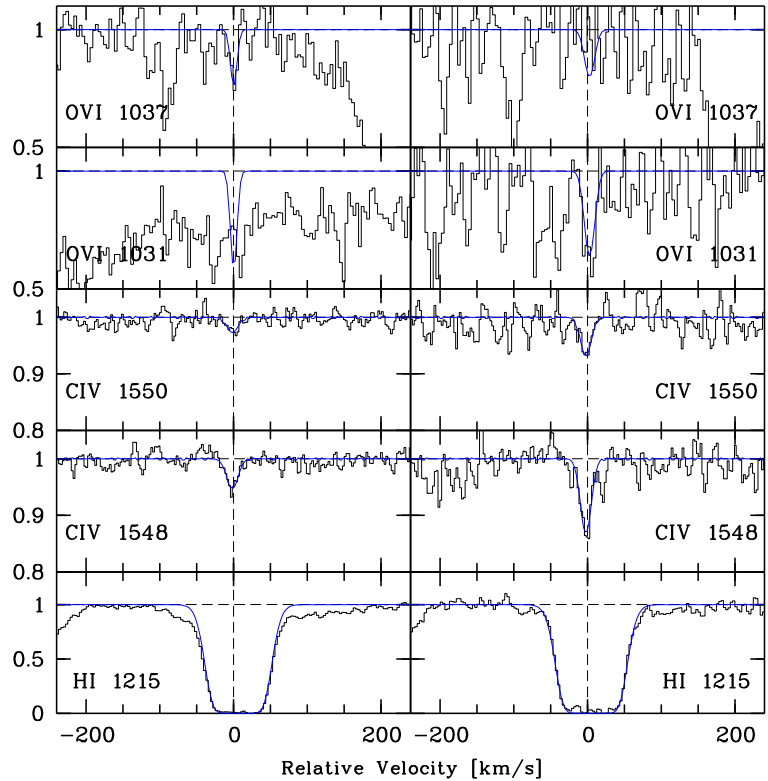}
   \caption{Class II system in \he. Panels and symbols as for
   Fig.~\ref{fig_spec0}, but for 
   $z=2.158112$.}  
   \label{fig_spec2} 
    \end{figure*}

   \begin{figure*}
   \centering
   \includegraphics[width=12cm,clip]{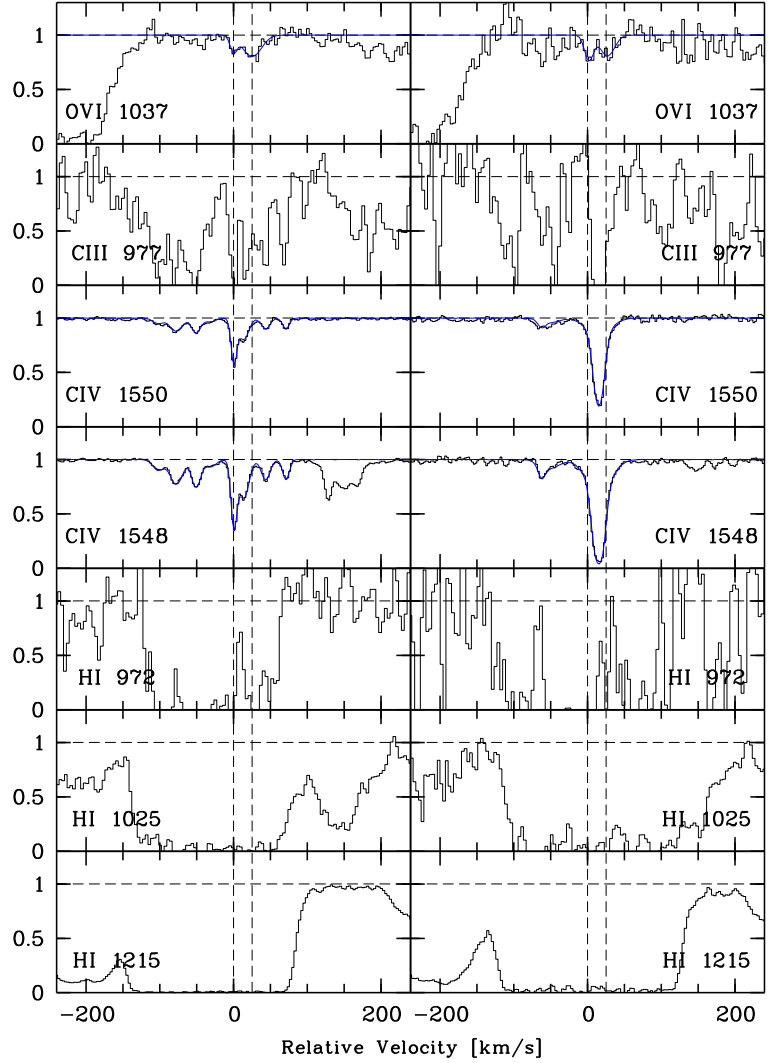}
   \caption{Class II system in  \he. Panels and symbols as for
   Fig.~\ref{fig_spec0}, but for    
   $z=2.200800$ and $z=2.201069$.} 
   \label{fig_spec3} 
    \end{figure*}

   \begin{figure*}
   \centering
   \includegraphics[width=12cm,clip]{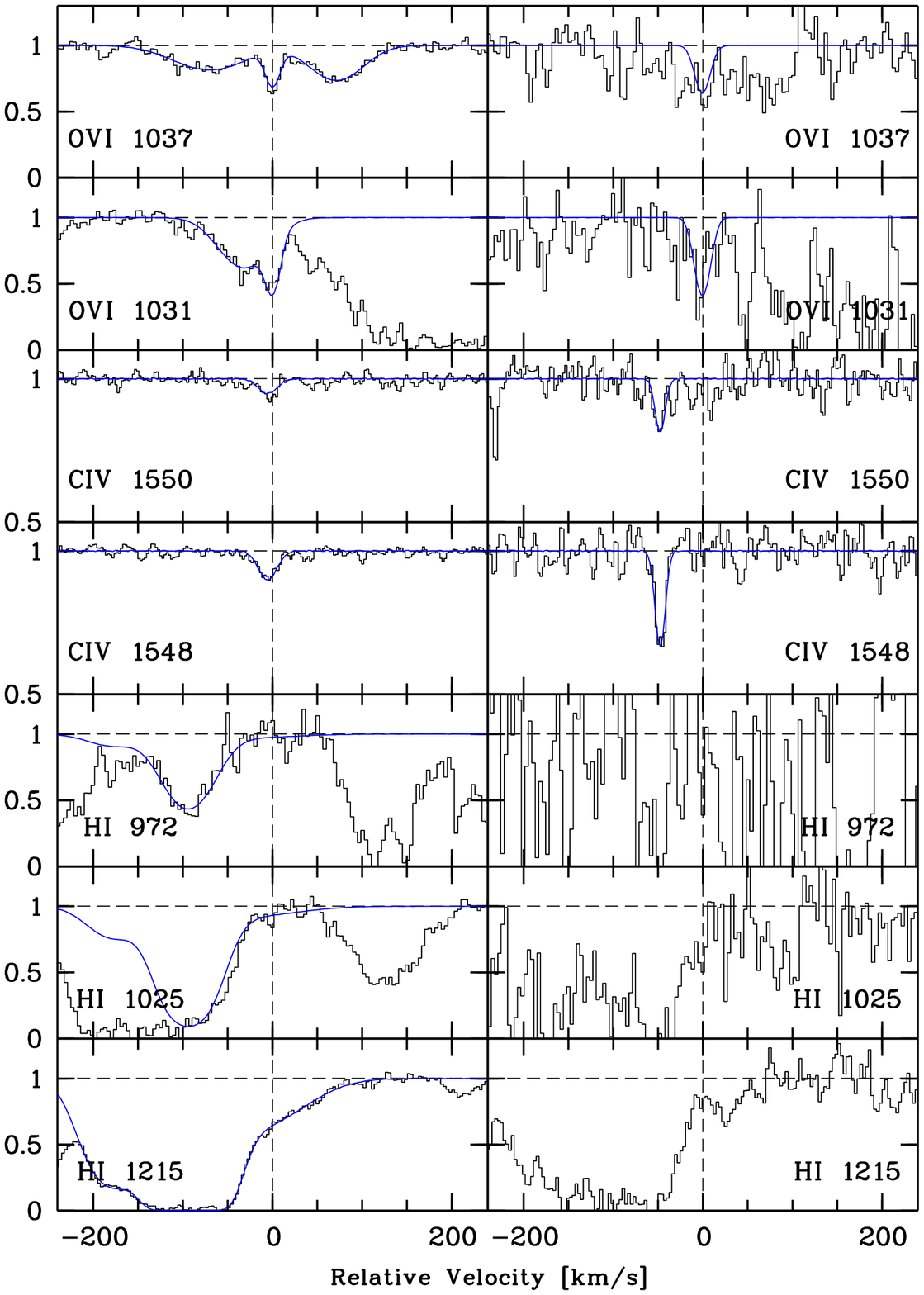}
   \caption{Class I system in \rx.  Panels and symbols as for
   Fig.~\ref{fig_spec0}, but for    
   $z=2.437271$.} 
   \label{fig_spec4}  
    \end{figure*}


\section{Absorption  systems}
\label{sect_abs}

\subsection{Selection}

Since the main goal of this paper is to study LOS differences between
\os\ absorption systems and the spectra have disparate S/N
ratios, the selection of the systems must be made as objective and unbiased  as
possible. 
 
We have searched for \os\ systems based on three independent methods, which we
applied to each of the LOS {\it separately}:

\begin{enumerate}
\item
Identify all \lya\ lines with \nhi $\ga 10^{13.5}$ \icm\ (checking for \lyb\
and \lyc\ to avoid confusion with strong metal lines in the forest), then look
for \os\ at similar redshift within $\pm 200$ \kms. When an \os\
candidate was found, check whether doublet lines could be \lyb\
interlopers. Note that this \nhi\ cutoff includes systems for which \os\
normally has not been detected by previous surveys. One caveat of the method
is that it will fail to find highly ionized systems with no \ho.

\item
Identify \cf\ absorption systems and look for \os\ at similar redshift.

\item
Blindly search for \os\ features taking into account the doublet ratio and
possible forest blending. This method is limited to clean and high S/N parts
of the forest, but it is advantageous in finding potential broad and shallow
\os\ absorbers.  This search was performed down to $z>2.0$ (\he\ A), $z>2.1$ (\he\
B), $z>2.2$ (\rx\ A), and $z>2.5$ (\rx\ B), which correspond to $3\sigma$
detection limits for \os\ of $\approx 10^{13.5}$ \icm\ (see  Fig.~\ref{fig_detection}).

\end{enumerate}

We found twenty one candidate \os\ systems. These are listed in
Table~\ref{table_OVI}. We then classified these candidates (column ``class''
in the Table) according to the following criteria:

\begin{itemize}

\item Associated system: redshift is within $v < 3\,000$ \kms\ of the
  systemic QSO redshift.

\item Class I: in both spectra both \os\ doublet lines are detected at the 
  $3\sigma$ significance level or better, and show the proper doublet
  ratios. Other metal species are present and \nhi $> 10^{13.5}$ \icm.

\item Class II: like Class I, but one of the \os\ doublet lines is detected at
  low significance in either of the spectra, or other metal species are not
  detected.

\item Class III: Like Class II but no metals are present and \nhi $<
  10^{13.5}$ \icm.

\end{itemize}

All these candidate systems are shown in Figures~\ref{fig_ClassIII} to
\ref{fig_spec7} and \ref{fig_spec8} to \ref{fig_spec10}.  
Associated systems
are treated  in the Appendix.

The class III absorbers represent the less certain identifications, all of
which are shown in Fig.~\ref{fig_ClassIII}.
Almost all of these were
discovered by method 3, except the  $z=2.418109$ system toward \rx. This
candidate does show \ho\ 
within $\pm 200$ \kms, but the 1037 line is also identified with 1031 of the
Class I system at $z=2.437271$, making the former identification uncertain.
For the rest of Class III systems, and when the S/N is high enough, we note
that an absorption feature at the position of \os\ appears always in the two
spectra. However, let us emphasize that in this sub-sample there is no clear,
unblended case of \os-only system. Consequently, the rest of this paper shall
deal only with the { ten remaining} Class I and II systems.

\subsection{Sample completeness}

One check of sample completeness is to  compare the number of selected systems
with previous studies. To this end we count systems that are class I or
II (since previous studies have considered systems with \ho).  For a total
redshift path surveyed of $\Delta z=0.649$ (based on \os\ detectability in the
lower S/N spectra of LOS B, which is one of the selection criteria),  we
find 7 systems with $W_{\rm rest} \geq$ 25\,m\AA. The estimated number
density of \os\ systems in our sample is thus ${d{\cal N}/dz} \approx 10.8\pm
4.1$ at $<z>=2.3$. This density is consistent with previous estimates by
 e.g., Carswell, Schaye \& Kim (2002; ${d{\cal N}/dz} \approx 11.4\pm4.0$)
or Reimers et al. (2001), so our particular selection criteria do not seem
to lead to a biased sample.

\subsection{Line parameters}

We used the FITLYMAN package in MIDAS to fit Voigt profiles\footnote{VPFIT was
tried independently on several systems, yielding exactly the same results.} to
class I and  II systems. When not otherwise stated, the \os, \cf\ and \ho\
lines were fitted independently so that identical velocity is not a
pre-requisite (except in Sect.~\ref{sect_photo} when we test
photoionization models). Allowing the components' velocity to be a free
parameter helps quantify the extent of these differences along and across the
LOS.  The \nhi\ was constrained only in a few systems.  In the much more
modest S/N spectra of the B QSOs we attempted a fit only to \os\ and \cf\
lines.

 In Figures~\ref{fig_spec0} to \ref{fig_spec7} we show transitions for
which we were able to obtain line constraints. The smoothed lines are the
results of the fits and their parameters are listed in
Table~\ref{table_AB}. The vertical dashed lines in the Figures indicate in
both panels the redshift of \os\ along LOS A  (except in
Fig.~\ref{fig_spec1a}). The transverse (proper) distance between LOS, $S(z)$,
was calculated using Eq. (1) of Smette et al. (1992) and Eqs. (7) and (8) of
Tzanavaris \& Carswell (2003), and assuming QSO angular separations and lens
redshifts $(\theta;z_{\rm lens}) = (3\farcs 20;0.73)$ and $(3\farcs 06;0.77)$,
respectively for \he\ (Lidman et al. 2000) and \rx\ (Kneib, Cohen \& Hjorth
2000).


\begin{table*}
\begin{minipage}[h]{18cm}
\caption{Intervening \os\ systems toward \he\ A and B, and \rx\ A and B}\label{table_AB}      
\begin{tabular}{lccccccccc}
\hline\hline   
       &\multicolumn{3}{c}{LOS A}        && \multicolumn{3}{c}{LOS B}\\
       \cline{2-4} \cline{6-8}
 Ion   & $z$      &$\log N$   & $b$      && $z$      &$\log N$   & $b$ &Class&QSO\\ 
\hline
 {\ho} & 2.010370 &14.39(0.01)& 41.3(0.9)&& &    & &II&\he\\ 
 {\cf} & 2.010531 &11.84(0.07)&  5.5(1.6)&&2.010452 &12.88(0.01)&7.6(0.3)\\
 {\os} & 2.010473 &13.39(0.21)&  9.8(6.1)&& &$<$13.7&\\
\hline
 {\ho} &          &           &          && 2.052476 &13.48(0.02)&24.5(1.6)&II&\he\\ 
 {\cf} & 2.053196 &11.77(0.09)&  5.2(2.1)&& 2.052499 &12.84(0.02)&8.9(0.5)\\
 {\os} & 2.052481 &14.01(0.06)&  8.3(1.7)&& 2.052440 &13.77(0.16)&7.4(4.5)\\
\hline
 {\ho} & 2.086492 &$>$15.11(0.01)& 29.9(0.7)&& &    & &II&\he\\ 
 {\cf} & 2.086522 &12.19(0.04)&  9.6(1.1)&& &$<$12.0&\\
 {\os} & 2.086682 &13.21(0.11)& 10.5(3.5)&& &$<$13.5&\\
\hline
 {\ho} & 2.158179 &$>$14.70(0.02)& 26.7(0.4)&& 2.158165 &$>$14.80(0.05)& 27.9(0.6)&II&\he\\ 
 {\cf} & 2.158093 &12.13(0.04)& 10.2(1.4)&& 2.158087 &12.52(0.03)&  9.5(1.0)\\
 {\os} & 2.158112 &13.22(0.11)&  4.6(2.9)&& 2.158139 &13.32(0.09)&  8.6(2.0)\\
\hline
 {\ho} &          &16.67(0.12)\footnote{From Lopez et al. (1999)}  &
       &&          &17.00(0.08)$^a$&&II&\he\\ 
 {\cf} &          &13.67(0.03)\footnote{Column density integrated over all
       fitted velocity components.}  
  &          &&          &13.92(0.03)$^b$\\
 {\os} & 2.200800 &13.00(0.20)&  4.8(3.3)&& 2.200822 &13.34(0.15)&  7.2(3.5)\\
 {\os} & 2.201069 &13.57(0.08)& 16.7(4.4)&& 2.201088 &13.44(0.15)& 12.6(5.9)\\
\hline
 {\ho} & 2.435220 &13.93(0.04)& 37.6(2.9)&&&&&I&\rx \\
 {\ho} & 2.436189 &14.86(0.02)& 36.6(1.0) \\
 {\ho} & 2.437021 &13.64(0.07)& 71.1(5.8) \\
 {\cf} & 2.437208 &12.54(0.03)& 13.9(1.2)&& 2.436724 &12.86(0.04)&  6.4(1.0)\\
 {\os} & 2.437271 &13.55(0.04)& 10.0(1.0)&& 2.437265 &13.70(0.10)& 10.6(3.7)\\
\hline
 {\ho} & 2.516735 &14.82(0.03)& 59.8(2.2)&&&&&I&\rx\\
 {\ct} & 2.516871 &12.18(0.20)& 26.7(0.0) \\
 {\cf} & 2.516871 &12.67(0.14)& 26.7(5.2) \\
 {\os} & 2.516735 &13.62(0.10)& 12.8(2.7)&& & & \\
 {\ho} & 2.517045 &14.88(0.03)& 31.9(1.3)\\
 {\ct} & 2.516995 &12.50(0.07)& 12.7(0.0)&&&$<12.53$ \\
{\cf} & 2.516996 &12.73(0.12)&12.7(1.6)&&2.516916 &12.92(0.06)$^b$& 17.6(2.6)\\
{\os} & 2.517045 &13.91(0.04)&16.7(1.5)&&2.516961& 14.06(0.07)$^b$& 22.8(4.2)\\
\hline
 {\ho} & 2.624316 &13.84(0.77)& 27.1(3.5)&&&&&II&\rx \\
 {\ho} & 2.624381 &14.08(0.44)& 23.8(2.1)&&&&&   & \\
 {\os} & 2.623886 &12.85(0.06)&  7.7(1.6)&& 2.623860& 13.08(0.17)&  1.9(0.9)\\
\hline
 {\ho} & 2.626526 &14.49(0.01)& 44.3(0.7)&&&&&II&\rx \\
 {\os} & 2.626749 &13.18(0.04)& 12.4(1.5)&& 2.626694& 13.18(0.14)&  9.6(4.7)\\
\hline
 {\ho} & 2.631842 &15.66(0.07)& 10.0(2.3)&&&&&I&\rx \\
 {\os} & 2.631480 &13.66(0.06)& 17.3(2.5) \\ 
 {\ho} & 2.632487 &15.98(0.06)& 21.6(3.4) \\
 {\os} & 2.631768 &13.54(0.09)&  8.9(1.2) \\
 {\os} & 2.632020 &13.83(0.05)& 12.9(1.1) \\
 {\ho} & 2.633119 &16.25(0.04)& 21.5(2.4) \\
 {\os} & 2.632495 &13.91(0.15)&47.3(13.6) \\
 {\os} & 2.632770 &14.43(0.04)& 52.5(4.0) \\
 {\ho} & 2.633998 &15.66(0.08)& 16.4(3.4) \\
 {\os} & 2.633161 &13.26(0.07)&  6.7(1.1) \\
 {\sus}& 2.633173 &12.74(0.06)&  8.5(1.5) \\
 {\os} & 2.633486 &13.03(0.16)&  7.8(2.1) \\ 
 {\os} & 2.633766 &13.40(0.05)& 10.1(1.2) \\
 {\os} &2.634119  & 13.18(0.05)& 12.7(1.6)&&    & 14.66(0.07)\footnote{AODM value.}

&\\
 {\sif}&          &13.88(0.02)$^b$&      &&    & 14.18(0.40)$^b$
  & \\

\hline
\end{tabular}
\end{minipage}
\end{table*}


\subsection{Description of Class I and II systems}
\label{intervening}

\paragraph{$z=2.010473$, $S=1.8$ kpc, Class II system in \he, see Fig.  
\ref{fig_spec0}}

This system was identified through the presence of \cf\ in both spectra,
  and absorption features at the position of both \os\ doublet lines. The
  detection of \os\ toward QSO A is based on the 1037 line, albeit at low
  significance level. At the position of \os\ in B the spectrum is too noisy to
  attempt any line-profile fit.  \cf\ can be fitted in both spectra and shows
  an order of magnitude variation in column density. \ho\ is constrained in A
  by \lya\ and \lyb.

\paragraph{$z=2.052440$, $S=1.5$ kpc, Class II system in \he, see Fig.  
\ref{fig_spec1a}}

For this \os\ system unfortunately only the 1037 lines are available. In
the B spectrum, this line matches \cf\ and a weak (\nhi $=10^{14.5}$) \ho\
line. \ct\ is not covered. Surprisingly at such small LOS separations, the
\ho\ profile is quite different in the A spectrum: no \ho\ is observed at the
position of \os\ and strong \ho\ lying redwards has no counterpart in the B
spectrum. There is a marginal ($3\sigma$ in 1548) detection of \cf\ toward QSO
A at $\approx 80$ \kms\ that matches the strong \ho\ system.

\paragraph{$z=2.086682$, $S=1.3$ kpc, Class II system in \he, see Fig.  
\ref{fig_spec1}}

In this system the detection of \os\ toward QSO A is
based on both doublet lines albeit at low significance level. 
\cf\ is clearly detected but blueshifted by $\sim 14$ \kms.
\ho\ is constrained in A by \lya\ and \lyb\ but both lines are saturated so we
only get a lower limit on \nhi. In B, from a non-detection of \cf, we
get a $5\sigma$ limit $\log N(\cf)<12.0$ \icm. At the position of \os\ the
spectrum is too noisy to attempt any line-profile fit.

\paragraph{$z=2.158112$, $S=0.9$ kpc, Class II system in \he, see Fig.
\ref{fig_spec2}}

For this system the detection of \os\ is based on the 1037 line in A and the
1031 one in B (although the 1031 line in the latter is only a $3 
\sigma$ detection). In
spectrum A there is a damped \lya\ system at $z=1.66$ that affects the
continuum at the position of the 1031 line. \cf\ is clearly detected in both
LOS.  The redshifts are almost identical ($\Delta v \sim 2$ \kms\ according
to the line centroids). \ho\ is constrained in A and B only by saturated \lya,
so the \nhi\ estimates have to be taken as lower limits.

\paragraph{$z=2.200800$, $S=0.2$ kpc, Class II system in \he, see Fig.
\ref{fig_spec3}}

This is a Lyman-limit system (LLS) with rich structure in \ct\ and \cf\ (also
in \sif, not shown in the Fig.). We note in the A spectrum that \ct\ and
\cf\ seem to follow each other, indicating that they arise in the same gas
volume. A similar behavior is seen in B, although with some additional
blending from interlopers.  Note the $v\approx 30$ \kms\ offset between \ho\
in A and B. The detection of \os\ is based solely on the 1037 line in A
because of \lya\ forest blending. The system was included because in both LOS
the redshift match with \cf\ is good and the small Doppler parameters indicate
metal lines.

\paragraph{$z=2.437271$, $S=1.4$ kpc, Class I system in \rx, see Fig.
\ref{fig_spec4}}

In this system the two \os\ doublet lines are clearly present in the A
spectrum. Although blended with shallow \lya\ (which have been included in the
line fits), the doublet ratios are consistent with a positive
identification. There is a good match in velocity between \cf\ and \os.  \ct\
is unfortunately blended and therefore not constrained.  From non-detection
of \nf\ we obtain the $3\sigma$ limit $\log N(\nf)<12.2$ \icm. For \ho, we have
considered three components in the fit which yields reliable \ho\ column
densities given that the non-saturated Ly$\gamma$
is available. The strong \ho\ velocity component closest
to \os\ appears blue-shifted by $\sim 100$ \kms. The component at $v=0$ \kms,
on the other hand, is weak and shallow, and has the largest Doppler parameter
in the sample, $b=71$ \kms.

 In B, there is a clear \cf\ system, which appears blueshifted by $43$ \kms\
with respect to \cf\ (and \os) in A, although we cannot discard that
undetected weak \cf\ lies exactly at the same redshift as in A.  From the
\lya\ profile of B (\lyb\ does not constrain \ho\ and \lyc\ is too noisy) we
observe that \ho\ does not vary between A and B, except for the flux excess at
$v=0$ \kms\ in LOS B.  The presence of \os\ in B is not conclusive;
nevertheless a 1-component fit gives same column density as in A within the
formal errors, and an almost identical $b$-value. Inclusion of \lya\ blends in
the fit was regarded as too complicated, given the noise level.

   \begin{figure*}
   \centering
   \includegraphics[width=12cm,clip]{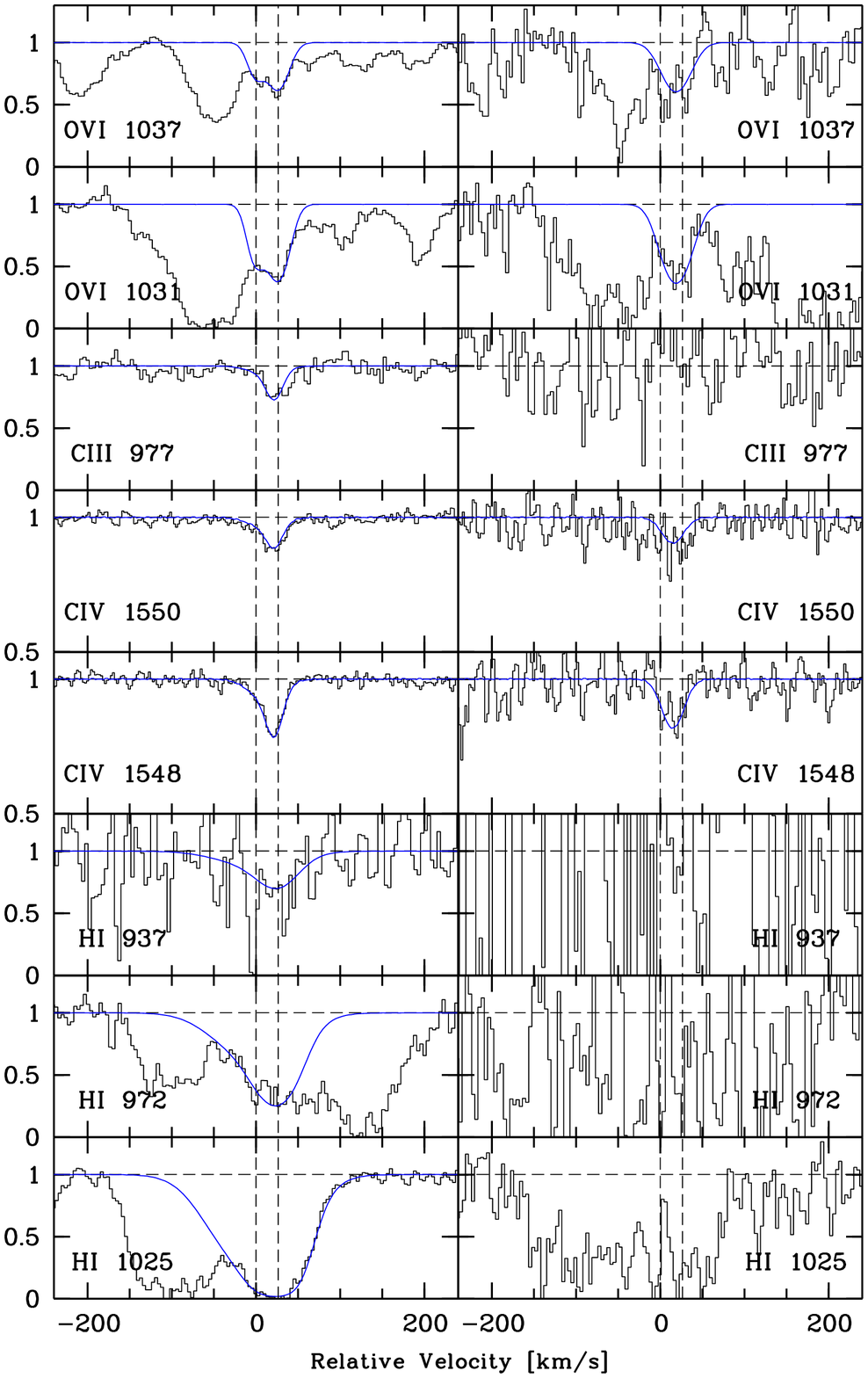}
   \caption{Class I system in  \rx.  
Panels and symbols as for Fig.~\ref{fig_spec0}, but for   
   $z=2.516735$ and $z=2.517045$.
}   
   \label{fig_spec5}  
    \end{figure*}
   \begin{figure*}
   \centering
   \includegraphics[width=12cm,clip]{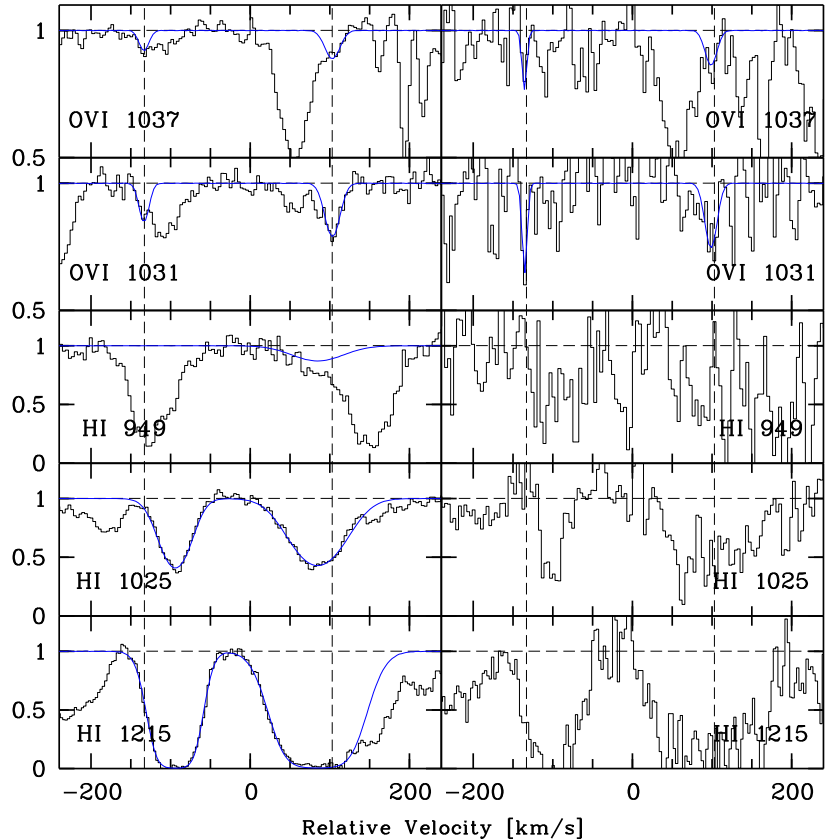}
   \caption{Class II systems in  \rx. 
Panels and symbols as for Fig.~\ref{fig_spec0}, but for   
   $z=2.623886$ and $z=2.626749$.
}   
   \label{fig_spec6}  
    \end{figure*}
   \begin{figure*}
   \centering
   \includegraphics[width=12cm,clip]{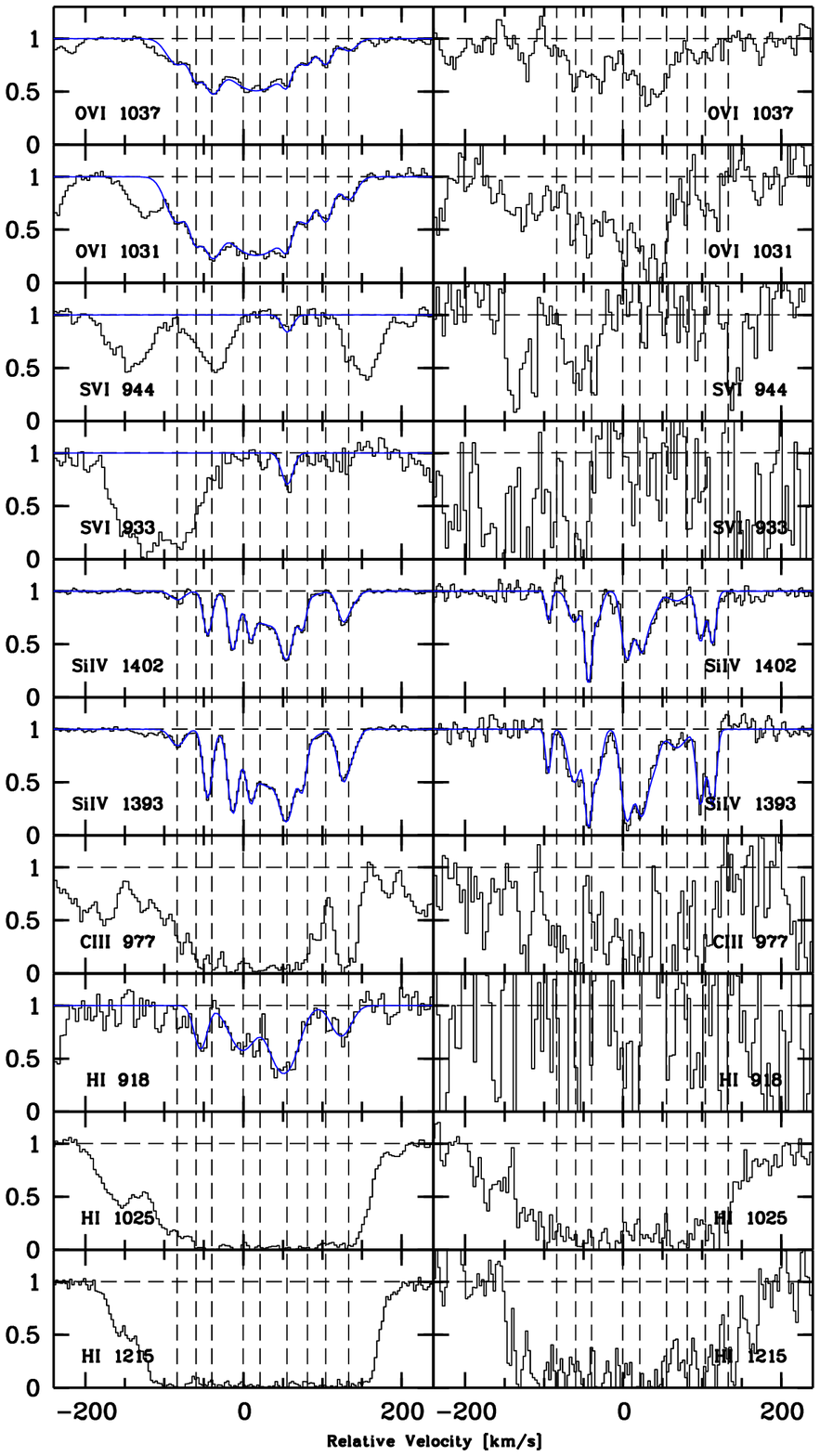}
   \caption{Class I system in  \rx. 
Panels and symbols as for Fig.~\ref{fig_spec0}, but for      
$z=2.631480$, $z=2.631768$, $z=2.632020$, $z=2.632495$, $z=2.632770$,
$z=2.633161$, $z=2.633486$, $z=2.633766$, and $z=2.634119$.
}   
   \label{fig_spec7}  
    \end{figure*}

\paragraph{$z=2.516735$ and $z=2.517045$, $S=1.0$ kpc, Class I system in \rx,
  see Fig. \ref{fig_spec5}} 

This \os\ candidate also shows the correct line-strength ratio. 
We fit two velocity
components in A, and note that \cf\ has also two components. The \ho\ was
constrained by a 2-component fit to \lyb\ through \lye, with redshifts fixed
at the \os\ values. \ct\ is also detected, which makes this system an
excellent case for ionization diagnostics.  From the
non-detection of \sif\ we get
the $3\sigma$ limit $\log N(\sif)<11.3$ \icm; from the
non-detection of \nth\
$\lambda 989$ we get the $3\sigma$ limit $\log N(\nth)<12.4$ \icm; and from
non-detection of \nf\ we get the $3\sigma$ limit $\log N(\nf)<12.0$ \icm.

In B there is absorption at the expected position of \os\ but a 2-component
fit here proved less reliable due to the lower S/N. We fitted a 1-component
profile and obtained the same value within errors as the total \os\ in A.  The
same procedure was applied to \cf. Therefore, the $b$-values displayed in
Table~\ref{table_AB} have no physical meaning if this system indeed is
composed of 2 velocity components along LOS B too.  The \ho\ was too difficult
to constrain in B again because of the high noise level.

\paragraph{$z=2.623886$ and $z=2.626749$, $S=0.6$ kpc, Class II systems in
  \rx, see Fig. \ref{fig_spec6}} 

For these two systems, the predicted position of \cf\ lies within spectral
gaps. We do not detect \sif, and \ct, \nf\ and \sus\ are blended.  From
non-detection of \sif\ we get the $3\sigma$ limit $\log N(\sif)<11.3$ \icm.
This is not proof against the identifications (neither of these ions is
expected in an \os\ phase if the gas is photoionized), but it makes the \os\
identification less reliable. We note that \ho\ is not aligned with \os\ in
either system.   Although attempting Voigt profile fits in B may seem
daring, they again show that \os\ cannot differ significantly from one LOS to
the other in neither system.

\paragraph{$z=2.631480$ through $z=2.634119$, $S=0.5$ kpc, Class I system in
  \rx, see Fig. \ref{fig_spec7}} 

This is a LLS with a total neutral hydrogen column density in A of $\log
N($\ho$)=16.56$ \icm.  This value comes from the unsaturated \ho\ $\lambda
918$ line, for which we fit 4 velocity components. This system was evident
already in the discovery spectrum of Bade et al. (1997) and lies $v=11\,000$
\kms\ from the QSO systemic redshift, so we regard it as intervening. Although
\cf\ falls in the gap, the \sif\ is evident and rich in structure.  Also,
lower ionization species are detected (not shown in the Fig.) such as \ctwo,
\nth, and \sit. 

We seem to have clearly detected \sus\ aligned with the strongest \sif\
component. There have been few reports of intervening \sus\ in the
literature. Savage et al. (2005) have studied a system at $z=0.2$ with
\nhi\ $\approx 15.0$ \icm, that shows \os, \nee, \sus, and a wide range of
low-ionization species. They have shown that photoionization explains the
low-ionization species, including \sus, but is not able to reproduce the \os.

We have fitted 9 velocity components in each spectrum. In A, we note the
disparity between the line centroids of \os\ and \sif.  However, the
velocities of the \sif\ and \ho\ are more similar (although due to higher
thermal broadening we do not resolve as many lines in \ho\ as in \sif).

The much lower S/N at \os\ in spectrum B prevents a reasonable fit. To compare
both LOS we first calculated 
the \os\ column density in A using both line profile fitting and the Apparent
Optical Depth Method (AODM). The total \os\ column density from the fits was
$\log N(\os) = 14.76\pm 0.07$ \icm, while from the AODM we obtained $\log
N(\os) = 14.74\pm 0.01$ \icm\ (integrating in the range $[-150,160]$ \kms, and
averaging the $\lambda 1031$ and $\lambda 1037$ values).  The good match
between these independent estimates indicates that the lines are not heavily
saturated. We repeated the pixel sum within the same velocity
interval along LOS B and obtained $\log N(\os) = 14.66\pm 0.07$ \icm,
indicating that the differences in \os\ column density between A and B are
negligible. Differences in kinematical structure between A and B are also much
more evident in \sif\ than in \os.

\section{Comparisons across the lines-of-sight}
\label{comparisons}


\subsection{Column densities}

   \begin{figure}
     \centering
   \includegraphics[width=9cm,clip]{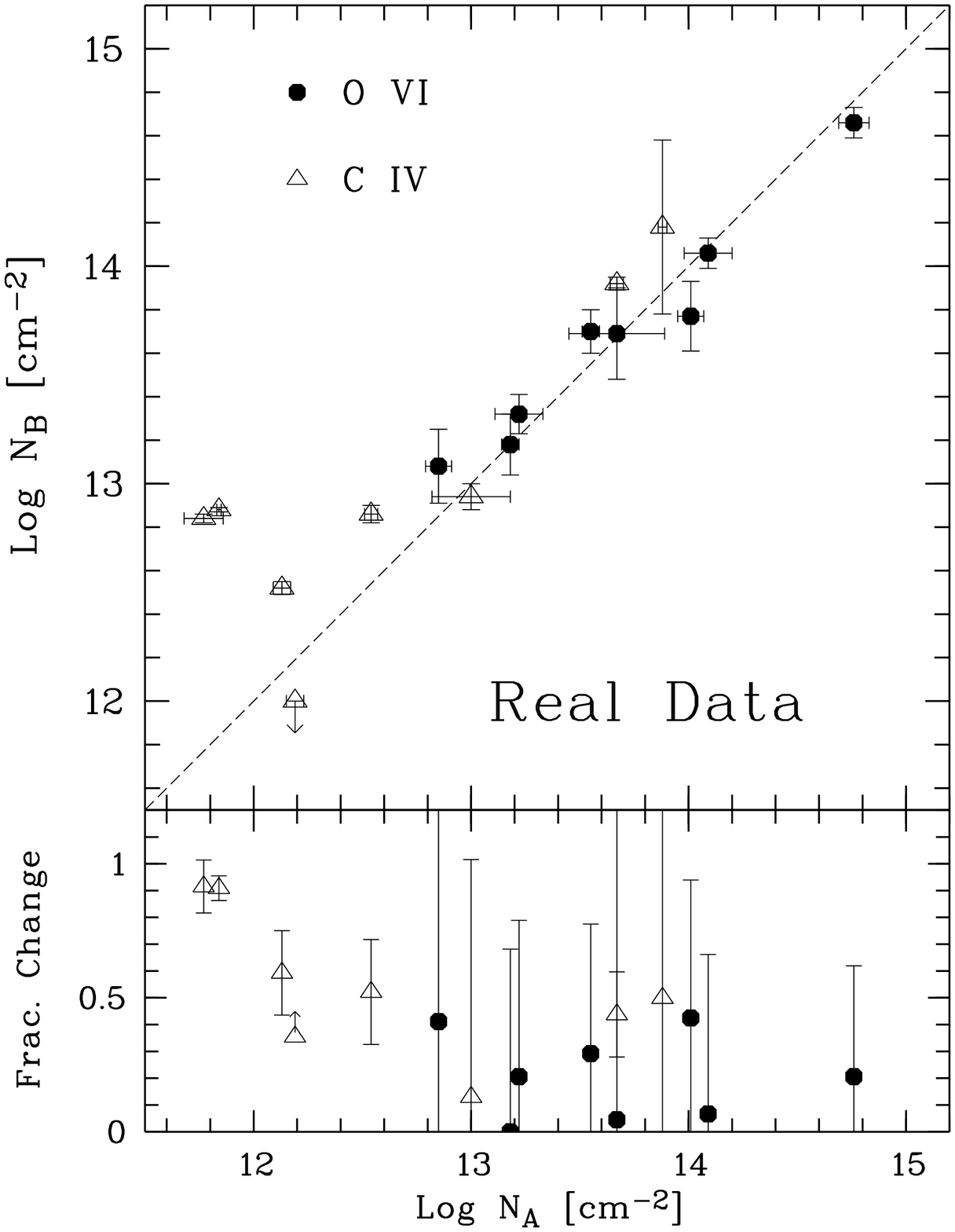}
   \caption{Comparison of column densities between the A and B lines of
   sight (upper panel), and fractional change between A and B, $|N_{\rm
   A}-N_{\rm 
   B}|/\max(N_{\rm A},N_{\rm B})$ for \os\ and \cf\ in \os\ systems. For
   clarity, the two upper limits on 
   $N(\os)_{\rm B}$ have not been displayed.  
   The associated systems have not been included.} \label{fig_colden}
    \end{figure}

   \begin{figure}
     \centering
   \includegraphics[width=9cm,clip]{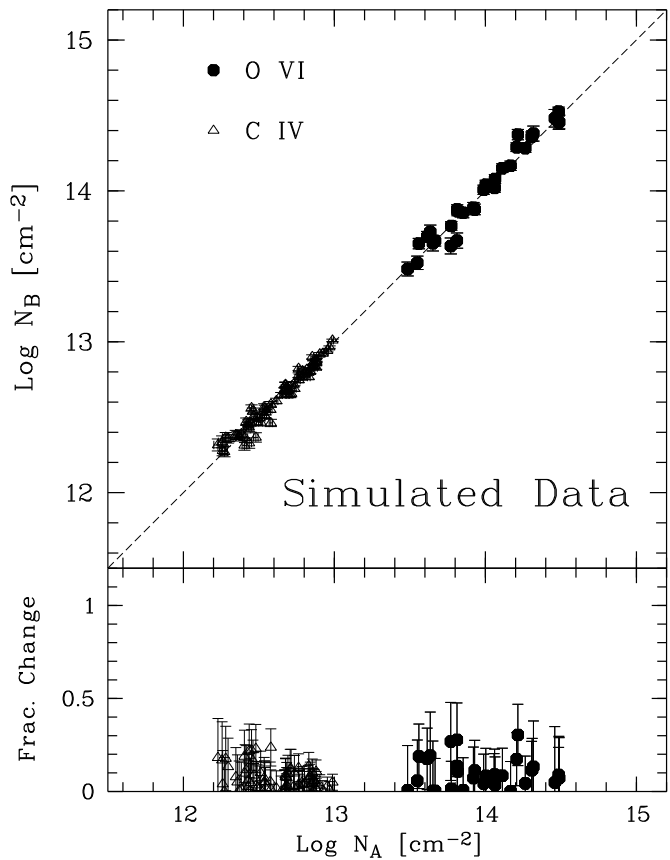}
   \caption{Same as in Fig.~\ref{fig_colden} but for simulated data. 
Column densities were recovered automatically from a sample of synthetic \os\
     and \cf\ lines with random redshifts and column densities and same S/N
     and blending levels as in the real data (see text for more details). The
     simulated column densities are identical in the two lines of sight.
} \label{fig_colden_MC}
    \end{figure}

   \begin{figure}
     \centering
   \includegraphics[width=9cm,clip]{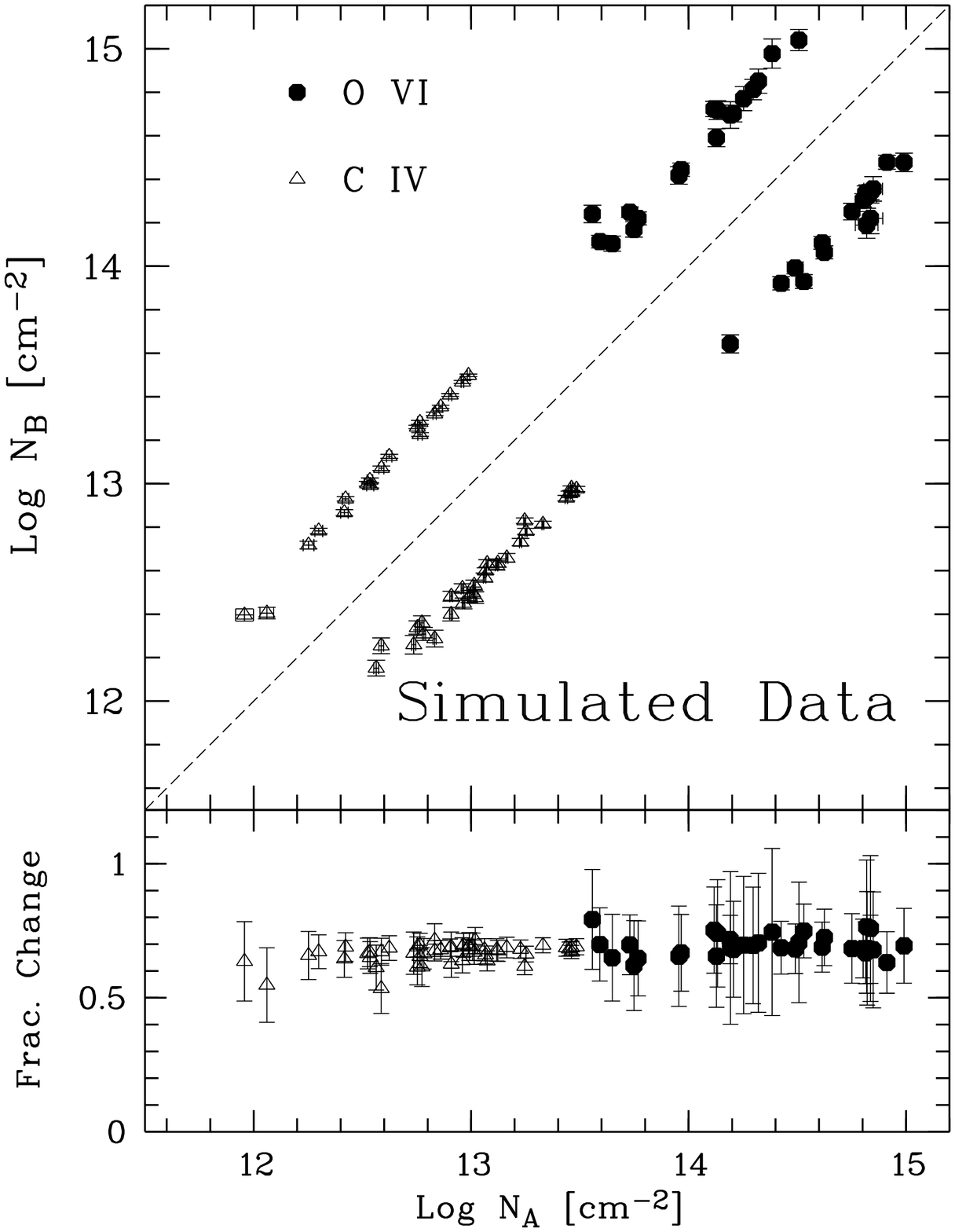}
   \caption{Same as in Fig.~\ref{fig_colden_MC} but the simulated column
densities differ by 0.5 dex (fractional change of $\sim$ 0.7), with the higher
column density present in either LOS A or B.  } \label{fig_colden_MC_diff}
    \end{figure}

Fig.~\ref{fig_colden} shows a comparison between column densities in A and B
according to the values listed in Table~\ref{table_AB} and excluding the
associated systems. { Column densities are summed over velocity components
when the other LOS has a value labeled with footnote $b$ in the Table.}  It
can be seen that the \os\ systems do not present significant LOS differences
(the 2 upper limits --not displayed in the Fig.-- are consistent with no
differences). In contrast 6 out of the 8 \cf\ systems significantly deviate
from the straight line of unity-slope (for the LLS at $z=2.63$ note that we
have used \sif\ as a proxy for \cf). This is also evident in the fractional
changes (lower panel of the Fig.) where we see that while the data is
consistent with a null change in \os, $6$ out of the $8$ \cf\ systems in which
we detect \os\ show, on the contrary, significant changes of up to 90 \%.  We
have to keep in mind that \cf\ column densities are measured redwards of the
forest and at higher S/N than \os\ column densities. { We estimate that
errors in \os\ column densities in the noisier B spectrum prevent us to assess
possible changes in \os\ of $\approx 40$ \% or lower.  }

 In order to assess the impact of blending and S/N in the \os\ region of the
spectrum compared with the \cf\ region, we have run Monte Carlo simulations.
Specifically, we synthesize \os\ and \cf\ lines with random redshifts and
column densities (within the observed ranges) and merge them with the real
data.  Two samples of spectra are created; in the first, the column densities
are identical in the two lines of sight, in the second we introduce a column
density difference of 0.5 dex (fractional change of $\sim$ 0.7).  In the
latter sample, the higher column density may be present in either LOS A or B.
For 300 realizations, we then attempt to recover the synthesized lines by
using the same search and fitting techniques as for the real data {\bf (the
only exception being that line widths are kept fixed at the simulated values,
which makes error bars on the column densities smaller than for the real
lines).  For the sample with equal column densities in the two LOS,
Fig.~\ref{fig_colden_MC} shows the recovered column densities in the same
format as in Fig.~\ref{fig_colden}. We find that the recovered \nos\ and
\nciv\ values in the synthesized systems have fractional changes $<0.2$ for
90\% and 95\% of the cases, respectively, i.e., consistent with equal column
densities in the two LOS.  In the sample for which the column densities differ
by 0.5 dex in the two LOS (Fig.~\ref{fig_colden_MC_diff}) the recovered
fractional change is on average $0.70$ for \os\ and $0.65$ for \cf, with a
rather small dispersion of $\approx 0.10$ for both samples. The fact that
column densities are recovered with high statistical significance for both
mock samples indicates that differences in S/N or effects of blending in the
\lya\ forest should not bias our results toward enhanced coherence in \os\
compared with \cf.}

{ It is also interesting to note that two
of the three 'Class I' systems, regarded as more certain, show respectively
low and high fractional changes for \os\ and \cf.  In summary, we 
conclude that we have detected an excess of transverse structure in the \cf\
profiles when compared with \os, a fact that lead us to conclude that a
significant fraction of \cf\ possibly does not trace the same volume as the
\os-bearing gas.}

At first glance, this result seems in contradiction with   previous
studies of \cf\ in lensed QSOs where coherence has been found on
scales considerably larger than probed here.  For example, in the compilation
of Ellison et al. (2004), there is only a difference of $\sim$ 30\%
for LOS separated by a few kpc.  The fractional column density difference
rises to 50\% for separations $\sim$ 100 kpc, whereas here we find 90\%
variation on a 1 kpc scale.  This apparent discrepancy can be explained
by the selection criteria of absorbers for our study.  Our fundamental
pre-requisite is that an absorber exhibit \os, whereas the majority
of \cf\ absorbers in previous samples do not show this species.
If we consider \cf-only (as opposed to \os+\cf) systems in our 2
lensed QSOs, we see a much higher level of coherence, with fractional
variations $\sim$ 10\% (Lopez et al., in preparation).

A second implication concerns the coherence length of the absorbers, i.e., the
scale at which significant changes in column density appear. Systems with \nos
$\ga 10^{12.9}$ \icm\ at $z=2.5$ must be much larger than the LOS separation
scales of $\approx$ kpc, given their lack of structure across the
LOS. Furthermore, in absorbers which exhibit both \cf\ and \os, the coherence
length of the former must be smaller than the latter.  Since potential \os\
systems occurring in just one LOS are likely not missing from our sample --
given the agreement of our \os\ number statistics with previous estimates --,
this last conclusion seems generic to all \os+\cf\ absorbers at high redshift.

It is interesting to compare our transverse LOS sizes with those
inferred from the ionization models of Simcoe et al. (2006).  For
$\sim$ 30 absorbers with multiple transitions, Simcoe et al. (2006)
derived metallicities and cloud sizes.  They concluded that the majority
of \os\ absorbers have high metallicities (up to $\sim 1/3 Z_{\odot}$) and
small sizes ($<$ 1 kpc) which inhabit volumes of 100 -- 200 kpc
around high redshift, star-forming galaxies.  Only around 25\%
of the \os\ absorbers modeled by Simcoe et al. (2006) had sizes
larger than 1 kpc.  However, Simcoe et al. (2006) also point
out that cross-section arguments indicate much larger dimensions;
combining this with the small cloud sizes yielded by their ionization
models led them to suggest a filamentary or sheet-like geometry.
In contrast to the mostly sub-kpc sizes inferred by Simcoe et al. (2006),
we find that all \os\ absorbers in our lensed QSOs (with LOS
separations $\sim$ 1 kpc) appear in both LOS and with very similar
column densities.  This either indicates \os\ clouds with sizes $>$ 1 kpc
or a high filling factor of small clouds.  The latter scenario was
proposed by Ellison et al. (2004) to explain the coherence between
Mg~II systems toward the triply imaged QSO APM08279+5255.  Although a Mg~II
absorber observed in one LOS was frequently observed in the others,
the difference in fractional column densities could be large, and the
line profiles were often very different.  However, this does not
seem to be the case for the \os\ absorbers.  In the cases where the
S/N is high enough that good Voigt profiles could be obtained
(e.g. Figures \ref{fig_spec3} and \ref{fig_spec5}), the velocity
profiles are very well-matched.  We therefore conclude that the
\os\ clouds included in this sample have sizes larger than $\sim$ 1 kpc.


   \begin{figure}
   \centering
   \includegraphics[width=9cm,clip]{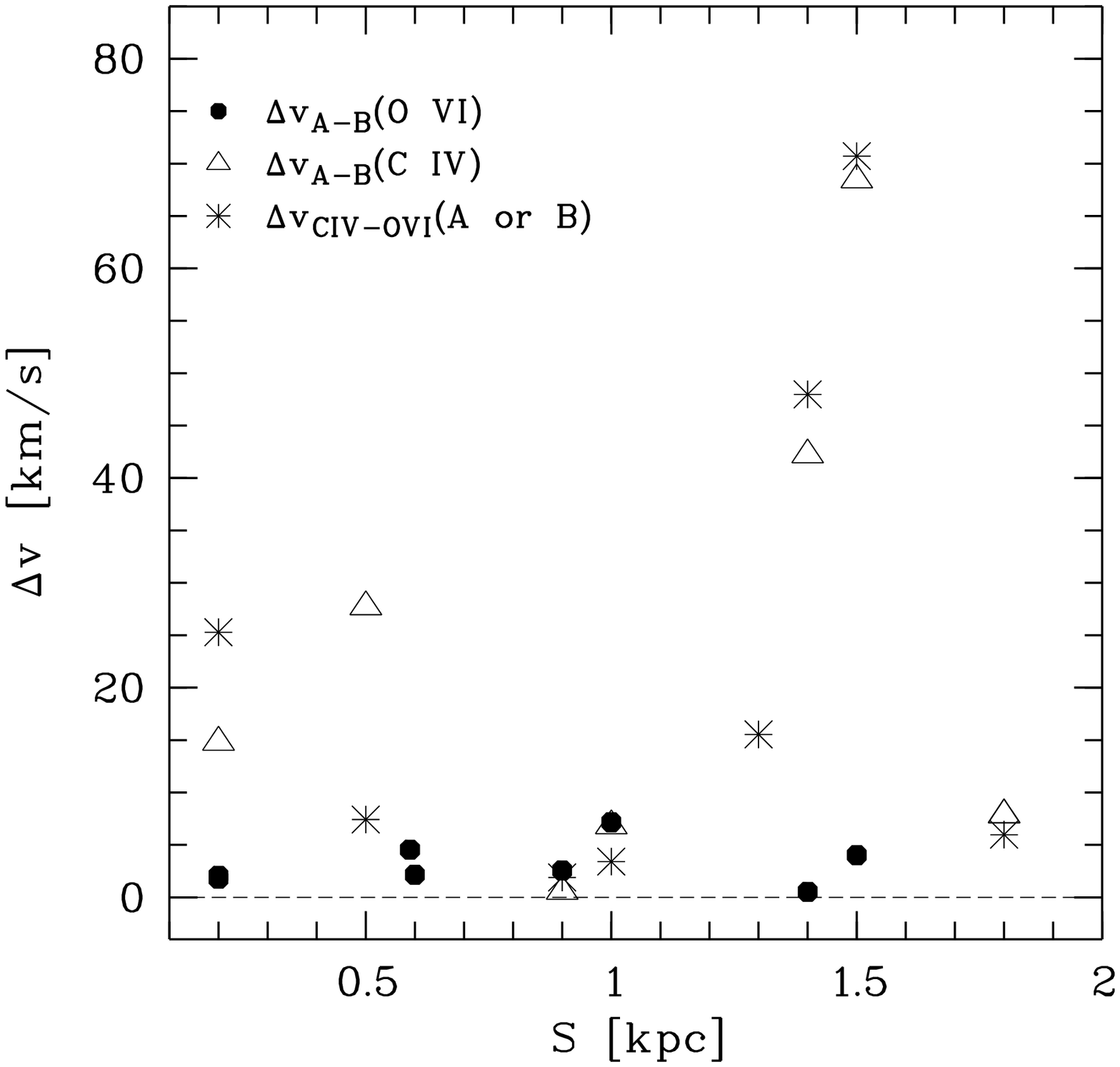}
   \caption{Velocity shear $\Delta v$ between LOS A and LOS B (filled circles
   for \os\ and open triangles for \cf) and between \cf\ and \os\ along LOS A
   or LOS B (asterisks) as a function of proper LOS separation, $S$. Errors in $\Delta
   v$ are 
   at most $\sim 3-5$ \kms\ (from line fitting).  The associated systems have
   not been included.} 
   \label{fig_vel}
    \end{figure}

\subsection{Velocity}

Fig.~\ref{fig_vel} shows the velocity shear between LOS A and B as traced by
\cf\ and \os.  We also show the velocity differences between \cf\ and \os\
along LOS A  or LOS B as a function of LOS separation. \ho\ was not
included in the Fig. because line centroids are not well constrained.  The
small number of systems in B for which we were able to get a Voigt profile
precludes a full quantitative analysis on velocity shifts. Nevertheless, we
can point out the following facts: (1) there is a clear offset between \os\
and \cf\ (or \sif) velocity components along LOS A in the three systems with
the highest \ho\ column densities, \nhi $>10^{15}$ \icm: the $z=2.086682$
absorber and the two LLSs. These systems also show differences in \cf\ or
\sif\ between LOS A and B ($\Delta v_{\rm A-B}(\cf)$), apparently independent
of LOS separation. Since LLSs are believed to be bound to collapsed
structures, $\Delta v_{\rm A-B}(\cf)\neq 0$ (though not necessarily $\Delta
v_{\rm \cf-\os}$(A)$\neq 0$) are expected on kpc scales.  (2) The 2 systems at
$z=2.052440$ and $z=2.437271$ (both probed at large separations)
show the largest offsets both in \cf\ between A and B, and between \cf\ and
\os\ in one of the LOS. Interestingly, these are the 2 systems in our sample
where \ho\ appears ``detached'' from the metals. 
These systems are considered in more detail below.

Altogether, velocity offsets reinforce the notion that the \os\ and \cf\
in our sample occupy (at least partially) different gas volumes.

\subsection{Line widths}
\label{line_widths}

Considering only the ``A'' spectra (with better S/N), in all 5 intervening
systems where $b$-values are relatively well constrained because they have
single velocity components, we find $b_\os<12$ \kms, implying temperatures
$T<1.4\times 10^{5}$ K. At this temperature and range of \nhi, models of
collisionally ionized plasma (e.g., Prochaska et al. 2004) predict much less
\os\ than what we observe here.  This indicates that photoionization is likely
to be the dominant process, as has been found for the majority of known \os\
systems at high redshift (Bergeron et al. 2002; Carswell, Schaye \& Kim 2002).

An important exception may be the LLS at $z=2.633$, some of whose velocity
components have $b$-values $\approx 40-50$ \kms. These may be artifacts due to
the more complex kinematics of this system (we note some degree of blending
with \ho\ interlopers), or a real hint of collisional ionization.

Regarding line-broadening mechanisms, previous samples contain a mix of
systems with lines dominated by either turbulence or thermal
broadening. Again, considering only the LOS A in our sample, out of the 5
intervening systems where $b$-values are well constrained, 4 show clear
thermal broadening, with $b(\os)\approx \frac{1}{4}b(\ho)$ within errors
($z=2.086682$, $z=2.158112$, $z=2.623886$, and $z=2.626749$), and one
($z=2.437271$) has anomalous $b$-values. The system at $z=2.517045$ has
$b$-values that suggest a mild contribution from turbulent-motions, but the
velocity components are not resolved, so $b$ is less reliable
there. Therefore, none of the systems along LOS A seems to be
turbulence-dominated.

Finally, if we compare LOS, we note that 4 out of the 5 systems where $b_\os$
could be measured in both spectra, present no transverse variations in the
line widths. In contrast, the (few) $b_\cf$-values measured are more disparate
between LOS. This is another hint that the \cf\ and \os\ absorbing gas
may occupy different locations.

\subsection{Comparison with \os\ in the Galaxy}
\label{Galactic}

We can compare the results of this Section with \os\ studies in the
Galaxy. The distribution of \os\ in the Halo is shown to be clumpy, with
fractional changes well in excess of what we observe at high redshift (Savage
et al. 2003); homogeneous \os\ is only shown by very small scales (a few pc;
Lehner \& Howk, 2004). But this is not the only quantitative difference with
our high-redshift observations. First, Savage et al. (2003) have indicated
that \cf\ might be kinematically less complex and less disturbed than \os\ in
the Halo; secondly, they find $N(\cf)/N(\os)=0.6\pm 0.1$ (average over $\sim
100$ LOS),
in contrast to all high-redshift systems studied in this paper, where we find
$N(\cf)/N(\os)\approx 0.1$ (excluding the LLS). Therefore, the systems probed
by these double LOS 
must be of different nature than the Galactic ones, a fact which should not
surprise given the different physics involved in each of them, warm-hot
photoionized and hot collisionally-ionized gas, respectively.


\section{Photoionization models and  a case study}
\label{sect_photo}

For the reasons outlined in Sect.~\ref{line_widths} in what follows we consider
the \os\ gas to be photoionized.  Photoionization balance is governed by 
the ionization
parameter, $U$, the ratio of hydrogen ionizing photon density to
total hydrogen density, \nh.  In principle, we could constrain $U$ by (a)
assuming an ionizing background spectrum, and (b) varying the gas metallicity
to match observed \nos\ and \nciv\ with model. However, this approach still
requires assumption of the [C/O] ratio and that the two measured
species arise in the same volume.  In addition, softer spectra are degenerate
with more diffuse (larger) clouds.  This degeneracy can be broken
with input from hydrodynamical simulations which correlate
 \nh\ with \nhi\ (with a mild dependence on temperature) at
least in the low-density regime and when hydrostatic equilibrium applies
(e.g., Schaye 2001).  However, in general, the lack of enough 
ionization stages of the same element makes detailed photoionization 
modeling uncertain.

One of the \os\ systems in our sample, at $z=2.5170$, is well suited for
testing photoionization because both \cf\ and \ct\ are present, so no
assumptions on relative abundances of different elements need be made. In
addition, \nhi\ is relatively well constrained by nonsaturated \lyc\ and
\lye. Therefore, we can use \ho, \ct\ and \cf\ as ``anchors'' for
photoionization modeling and investigate whether \os\ can be reproduced by
the models.

\subsection{A single-phase model}


   \begin{figure}
   \centering
   \includegraphics[width=9cm,clip]{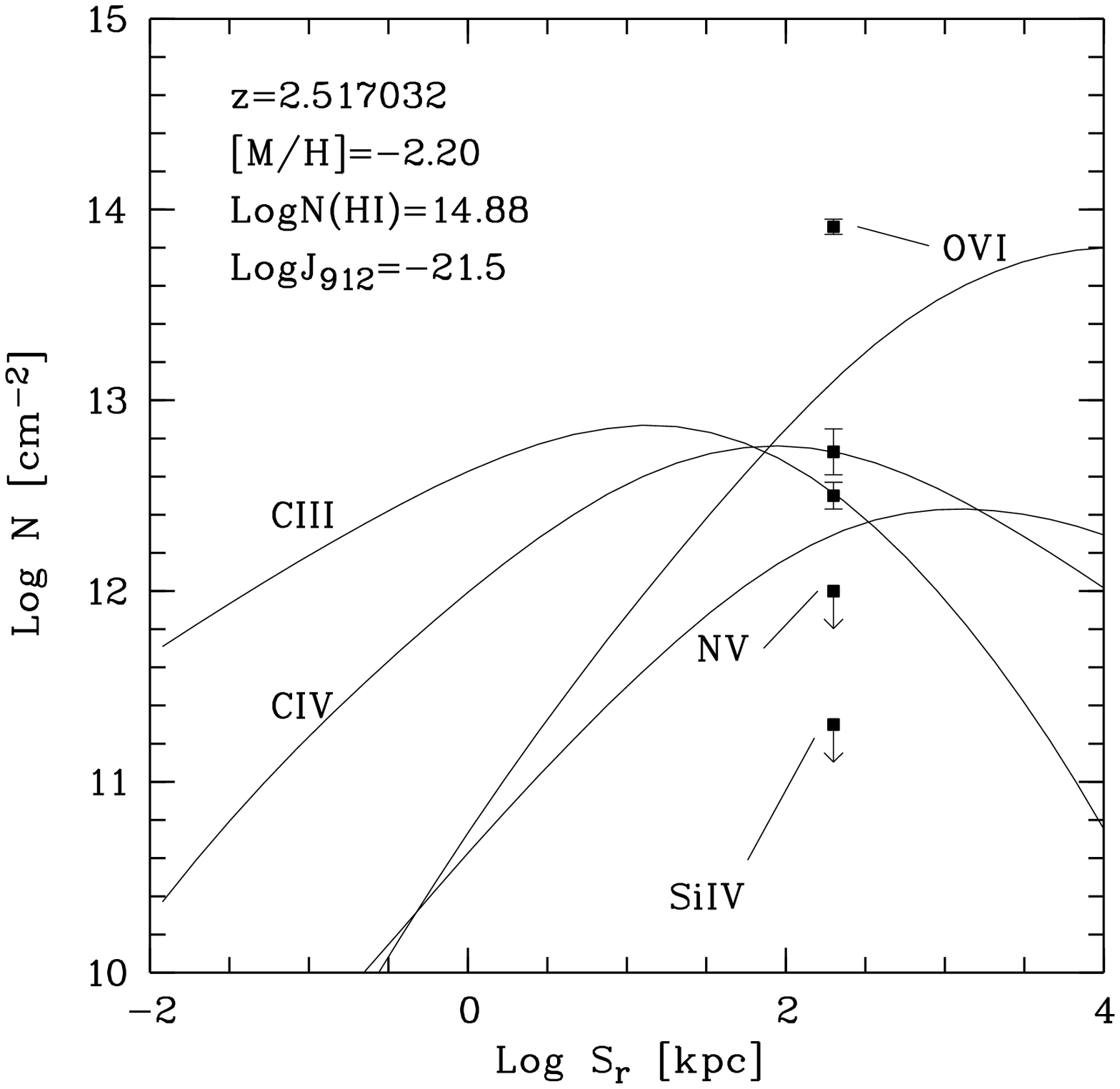}
   \caption{Single-phase model for \os\ system at $z=2.517032$ . Curves are
   theoretical estimates of column densities as a function of cloud radial
   dimension $S_r$ for each of the relevant species (labeled on top or below
   the 
   curves). The model assumes a metallicity [M/H] $=-2.20$ and solar relative
   abundances. The \sif\ curve lies below the $y$-axis range.  Observed values
   and their errors all are  displayed at the $x$-axis position that matches
   \ct\ and    \cf.  }  
   \label{fig_CLOUDY}
    \end{figure}

We start by asking whether photoionized gas can simultaneously host \ct, \cf,
and \os. To test this possibility we first need to re-fit the data with
tied velocities. The first column in table~\ref{tab_photo} lists the observed
column densities for a best-fit redshift of $z=2.517032$.  Note that these
differ only slightly from those ones in Table~\ref{table_AB}.  Assuming \ct\
and \cf\ coexist in the same volume, we build a grid of photoionization models
using the CLOUDY package (Ferland et al. 1993), with the geometry approximated
by a parallel-slab geometry. The ionizing agent is described by the Haardt \&
Madau (1996) spectrum of the UV background (UVB) at $z=2.5$ with normalization
$\log J_{912}=-21.5$. The spectrum includes contributions from QSOs and AGNs
only, whose spectra are described by power laws $F\propto \nu^{-\alpha}$ with
spectral index $\alpha=1.8$; inclusion of star-forming galaxies appears to
have little effect on the two carbon species treated here but may have an
effect on \os\ (see Fig. 3 in Simcoe, Sargent \& Rauch 2004, and
Sect.~\ref{uncertainties} below). Solar abundances
are taken from Grevesse \& Sauval (1998), except for oxygen which is taken
from Allende Prieto, Lambert \& Asplund (2001).

Fig.~\ref{fig_CLOUDY} shows model column densities for \ct\ and \cf, and
(assuming solar [C/N,O]) also for \nf, and \os, as a function of radial size
$S_r$. $S_r$ is proportional to $U$ through fixed $J_{912}$ and \nhi.  The
\ct/\cf\ ratio determines $U=10^{-0.8}$, which for the assumed value of
$J_{912}$ corresponds to $S_r=200$ kpc ({ note that these numbers appy to a
single cloud in this system}).  Having determined $U$, the carbon
abundance can be varied to match the observed \ct\ and \cf\ column
densities. The best match is found for [C/H] $=-2.20$.  This is somewhat over
the median but within the spread of \lya\ forest metallicities around [C/H]
$=-2.8$ found by Simcoe, Sargent \& Rauch (2004) also from photoionization
arguments.  However, it can be seen that \os\ is not reproduced by the
model unless oxygen is overabundant with respect to carbon.

With the assumed UVB intensity, the above conditions correspond to a gas
density \nh $=10^{-4.1}$ cm$^{-3}$.  The \nh-\nhi relation yields \nh $\la
10^{-3.8}$ cm$^{-3}$ (value assumes a kinetic temperature $T\la 1.0\times
10^5$ K from the newly-fitted $b_{\rm \cf}=12.0$ \kms). The CLOUDY solution
would match this independent value for a factor of 2 larger clouds, a factor
of 2 more intense UVB, or a combination of both.

\subsection{A multi-phase model}
\label{sect_multiphase}


\begin{table}
\begin{minipage}[t]{9cm}
\renewcommand{\footnoterule}{}
\caption{Two-phase photoionization model for cloud at $z=2.517032$}
\label{tab_photo}      
\centering                          
\begin{tabular}{l  r r r }        
\hline\hline                 

 & Observed& \multicolumn{2}{c}{Model}\\
        \cline{3-4}
   &      & Low & High\\    
\hline               
[C/H]         &      &     $-2.20$ &   $\la -2.10$  \\
$\log U$      &      &     $-1.15$ &   $ -0.65$     \\
$\log n$ [cm$^{-3}$]&&    $-3.75$  &   $ -4.3$     \\
Size [kpc]    &      &   20        &   $\la 250$ \\
{[C/O]}       &      &      0\footnote{Assumed.}      &   $\approx -0.8$    \\
{[C/N]}       &      &      ...    &   $\ga 0.2$    \\
$\log N(\ho)$ &14.88 &     14.58   &   14.58    \\
$\log N(\ct)$ &12.50 &     12.50   &   $\approx 12.00$    \\
$\log N(\cf)$ &12.73 &     12.43   &    12.43     \\
$\log N(\os)$ &13.92 &     12.21   &    13.91     \\
$\log N(\nf)$ &$<$12.00 &     11.60   &   $\la 12.00$      \\
\hline                                   
\end{tabular}
\end{minipage}
\end{table}

Assuming homogeneity, it is difficult to think of such large clouds hosting
\ct-\cf\ gas, and yet showing transverse structure on kpc scales, as observed
in Fig.~\ref{fig_colden}.  An overestimate of the size due to an incorrect UVB
normalization can be ruled out; uncertainties inherent to this number do
exist, but these do not reach two orders of magnitude.  A more plausible
situation might be that \cf\ arises in two distinct gas phases, which in turn
would explain the offsets observed between \cf\ and \os, since the absorption
would occur in different volumes.\footnote{Note that we refer to phases of
  photoionized gas. As we have argued in Sect.~\ref{line_widths}, we do not
  find evidence of collisionally ionized gas. }

We investigated photoionization models composed of the following two media: a
low-ionization phase in which most of the observed \ct\ but only a fraction of
the total \cf\ are present; and a high-ionization phase in which \cf\ and most
of the \os\ occurs.  Using the simplest approach, \ho\ is  distributed so
as to keep the same \cf/\ho\ ratio in each media. As for the \cf\ (and \ho)
fraction, very unbalanced fractions result in conditions that
either approximate to the single-phase model described above or produce
unrealistic parameters (sizes of several Mpc at [M/H] $=-1$ to reproduce
\os-only clouds).  Therefore, a ``50-50'' distribution of the observed \cf\
appears reasonably representative of the possible variety of cases. Running a
new grid of CLOUDY models this time using the modified column densities but
the same approach to find $U$ and $Z$, we determined conditions for both
phases independently.

The results for the  (newly-fitted) $z=2.517032$ system are displayed in
Table~\ref{tab_photo}.  Getting parameters for the low-ionization phase is
straightforward because we match the \cf/\ct\ ratio with the observations
without assumptions on relative abundances, as in the single-phase model. This
results in a cloud of radial size $S_r=20$ kpc at [C/H] $=-2.20$. $N(\os)$ is
50 
times lower than in the high-ionization phase (assuming solar [C/O] -- we lack
of lower ionization stages of oxygen to constrain [O/H] in this phase), too
low to be detected in our data and even for overabundant oxygen it would not
be detectable.  To model the high-ionization phase a [C/O] ratio has to be
assumed since we require that \ct\ does not contribute significantly to the
observed value. That ratio is constrained imposing that the radial size be
smaller than the maximum size allowed by the \ho\ line width, if it were
purely Hubble broadening, or $S_r<250$ kpc for this particular system. To match
the data, [M/H] $=-2.10$ and [C/O] $=-0.8$ is required (we discuss this
further in the next sections). In addition, our limit on \nf\ requires [C/N]
$\ga 0.2$ dex (in the low-ionization phase this ratio is also consistent with
the limit for \nth\ --see \ref{intervening}.).


\subsection{Characteristic sizes along the lines of sight}
\label{sect_sizes}

The two-phase model explains simultaneously
the observed column density differences in \cf, the velocity shifts between
\cf\ and \os\ and the homogeneity of \os\ on kpc scales. In fact, $S_r\approx
20$ kpc sizes, if also representative of transverse dimensions, perfectly
produce fractional changes in \cf\ of the order of what we observe here, as
demonstrated by size estimates by statistical means from observed fractional
changes in \cf\ on kpc scales (Rauch, Sargent \& Barlow 2001).  As for the
high-ionization phase, \os\ clouds that extend over a few hundred kpc are
expected to produce featureless absorption on the small scales probed with the
present data, and, also fitting well to our model, studies of wider QSO pairs
have  found no correlation of strong \cf\ systems over $\sim 300$ kpc scales 
(Becker et al. 2004). 


Our  ``50-50'' model is also consistent with the good match
between \cf\ and \ct\ line profiles in all systems where \ct\ is detected (the
2 LLSs, and $z=2.516$ systems); those cases show that the fraction of \cf\ in
the low-ionization phase cannot be arbitrarily low.

According to our model, if one LOS happens to pass through the two phases
while the other LOS does so only through the more extended one, then column
density differences of up to $0.3$ dex should arise in \cf, \ct, {\it
and} \ho, while almost the same \os\  should be measured 
along both LOS, a situation that is effectively observed in our sample.
Regarding \ho, although no firm constraints could be made on column densities,
there are some indications of variations of this level in the systems at
$z=2.158112$, and at $z=2.200800$.

Finally, note that we have imposed homogeneous \cf/\ho\ ratios, so our model
explains \cf\ gradients across the LOS as a consequence of the smaller size of
low-ionization phase only, not because of inhomogeneous ionization conditions
(within this phase). On the other hand, homogeneous ionization in the extended
\os-phase is supported by the similar line widths observed in LOS A and B.
Being thermally-dominated, these similar widths imply the absence of
significant temperature gradients, and thus ionization must be homogeneous (at
least on the kpc scale and with local ionizing sources being unimportant).

\subsection{Uncertainties}
\label{uncertainties}

The results from previous Section suffer from some model and measurement 
 uncertainties.

The first question one may raise is whether the \os\ systems are really
photoionized. In Sect.~\ref{line_widths} we have argued that \os\ line widths
of the general population in our sample suggest photoionized gas. In
particular for the $z=2.517032$ system, although its \os\ Doppler value is not
very well constrained and does not rule out collisional ionization (this
particular measurement is subject to systematics anyway due to the unresolved
velocity components), we note that both the presence of \ct\ and the narrow
\cf\ profile hint at photoionized gas (see Prochaska et al. 2004). 

A second concern is what effect errors in the continuum can have on the CLOUDY
parameters, specially given the limited S/N of our data in this part of the
forest. To explore this we recomputed column densities for the $z=2.517032$
system using flux levels that were displaced artificially by 1$\sigma$. The
new column densities differ by $\sim 0.08$ dex. The \ct/\cf\ ratio is
therefore determined with an uncertainty of at most $\sim 0.16$ dex, which
translates into a $\sim 0.20$ dex uncertainty in $\log U$. This means that a
wide range of possible radial sizes, $S_r \sim 100-500$ kpc for the
high-ionization and $S_r\sim 10-80$ kpc for the low-ionization phase, are
consistent with the data (of course this range assumes that $J_{912}$ is
accurate to this limit or better). Therefore, although the derived sizes are
quite sensitive to changes in column density, departures from the 50-50 model
still result in two size ranges that differ by one order of magnitude.

We then consider the shape and normalization of the UVB. As mentioned before,
inclusion of star-forming galaxies (Simcoe, Sargent \& Rauch 2004) does not
affect the region encompassing the \ct\ and \cf\ ionizing thresholds, but may
lower the UVB at the \ion{He}{ii} edge by $\sim 0.3$ dex.  This means that
radial sizes (derived from the \ct/\cf\ ratio) remain unaffected, but our
result on the C/O ratio may be affected at this level. Secondly, the UVB
models contain also uncertainties in the spectral index of the quasar spectra
(assumed to be pure power laws). The UVB hardness affects \cf\ and \os\
differentially (Schaye et al. 2003), thus directly affecting our results on
the C/O ratio. For instance, a spectral index of $\alpha=1.5$ (instead of the
assumed $\alpha=1.8$) favors a higher prediction of \os, and thus a lower C/O
by $\approx 0.4$ dex. Finally, the UVB normalization has been constrained with
independent methods (statistics of metal species and proximity effect) leading
to similar results, and is therefore subject to less important uncertainties.

In conclusion, besides the fact that this is just a case study and may not be
widely applicable to absorbers in general, the main caveat of our model seems
to be the shape of the UVB, affecting more the derived [C/O] than the sizes.

\section{The source of metals}
\label{sect_metals}

\subsection{Relative abundances}

Regardless of cloud size, important inferences on relative abundances can be
made. In the single-phase model depicted in Fig.~\ref{fig_CLOUDY} we see that
\os\ and \cf\ are not reproduced simultaneously unless [C/O]$ \approx -0.7$,
and a similar situation is found for the two-phase model (but let us
emphasize that the predicted [C/O] abundance is affected by the shape of the
UVB: a harder [flatter] spectrum, say with $\alpha=1.5$, would yield [C/O]
$\approx -0.4$ only).  As for  nitrogen, \nf\ should be detected according to
our CLOUDY models but it is not, implying [C/N]$\ga 0.2$.

Such abundances may conform to gas that has been enriched recently before the
  absorption occurred -- in contrast to the yields of an early pre-enrichment.
  The present abundance pattern is similar to direct measurements of C/H and
  O/H in a few damped \lya\ systems (DLA) at similar metallicity levels (e.g.,
  D'Odorico \& Molaro 2004). Furthermore, low [N/O] is found in a fraction of
  low-metallicity DLA, an abundance ratio that has been interpreted as an
  indicator of recent chemical enrichment (e.g., Pettini et al. 2002; 
  Prochaska et al. 2002).

Our result on [C/O] is also to be confronted with measurements of carbon and
  oxygen in Galactic metal-poor stars. For example, Akerman et al. (2004) have
  studied the metallicity dependence of [C/O] and confirmed the drop of [C/O]
  to $-0.5$ at [O/H] $\approx -1$ ([C/O] $=-0.7$ if non-LTE corrections are
  considered, see Fabbian et al. 2005). This is roughly the regime we are
  probing in the \os\ system studied in Sect.~\ref{sect_multiphase}. The
  absorbing gas would be enriched by a first generation of massive stars, just
  as predicted by Galactic chemical evolution models after completion of the
  first Gyr of the Halo formation.

Alternatively, the observed gas may be chemically young but have been
released to the IGM long-time before the absorption occurred (say, over 2 Gyr
before, $z>\sim 6$).  Though plausible, this scenario requires another
ingredient, namely that the gas that was further processed to reach higher C/O
in that time has not reached yet the physical volume probed by the $z=2$
observations.

It therefore seems that one viable way to reproduce the data on \os,
\nf, and \cf\ via photoionization is with CNO abundances from late rather than
early injection of metals. An extended phase of such enriched gas might be in
broad line with [C/Si] $\approx -0.8$ measured in the high-$z$ IGM and in
similar absorption systems but with a different technique (Aguirre et
al. 2004).

\subsection{\os\ outflows?}

The systems at $z=2.437271$ and $z=2.052440$ are perhaps good examples of
the two-phase scenario and deserve further analysis. Besides their particular
velocity offsets (between \cf -\os\ and \ho\ along LOS A, and between LOS A
and B in \cf\ --see Fig.~\ref{fig_spec4}) these systems also have the highest
$N(\os)/N(\ho)$ ratios in the sample, $1.25$ and $1.6$, respectively, so both
must be highly ionized. Systems with these $N(\os)/N(\ho)$ ratios are uncommon
but have been reported both at high (Reimers et al. 2001; Bergeron \&
Herbert-Fort 2006) and low redshift (Tripp, Savage \& Jenkins 2000).

The observed velocity offsets agree well with the multi-phase model, with a
fraction of \cf\ in a low-ionization phase that is kinematically detached from
a more extended \os\ phase. In the $z=2.437271$ system along LOS B, the
fraction of \cf\ that would arise in the high-ionization phase (expected at
the position of the dashed line) is lost in the noise. In addition, the \ho\
profile in B suggests a slightly lower column density than in A, just as
expected (Sect.~\ref{sect_sizes}) from our two-phase model.  A similar
  situation is found for the $z=2.052440$ system.

Repeating the CLOUDY simulations for these systems is more uncertain because
\ct\ is not available, and thus $U$ can only be constrained if a value for
[C/O] is assumed. Taking the $z=2.437271$ system along LOS A as an example,
the 50-50-\cf\ distribution yields a high-ionization phase with $S_r=50$ kpc
for [C,O/H] $=-0.75$ (i.e., assuming solar [C/O]).

This \cf\ gas, perturbed on kpc scales (Rauch, Sargent \& Barlow 2001) and two
orders of magnitude more metal-rich than the average \lya\ forest (Simcoe,
Sargent \& Rauch 2004), is highly suggestive of gas close to overdense regions
with a high star-formation rate. Moreover, the transverse velocity
offsets and column density differences of \cf\ and \ho\ also suggest possible
gas outflows with ejection velocities of a few hundred \kms, due to violent
dynamical processes driven by supernovae explosions.

Highly ionized, metal-rich \os\ systems detected toward single QSOs have been
interpreted as gas outflows before (e.g., Aracil et al. 2004; Simcoe et
al. 2006; Fox et al. 2007). The tens-of-\kms\ shifts we detect here in \cf\
might be a signature of such possible effects across the LOS (although we
stress that these two systems have \os\ line widths consistent with
photoionization and \nhi\ much lower than those probed by the sample of Simcoe
et al. 2006).  The observed profiles could alternatively be explained by gas
associated with high-redshift disks or extended halos, but recall from
Sect.~\ref{Galactic} that such halos would have to be quite different in
nature from the local one.

Another interesting point is that only \cf\ shows disturbance on these scales,
not \os. One possible interpretation is that this kind of outflows do not
transport highly ionized gas. Note that this picture does not depend on the
lower significance of the \os\ detection in the $z=2.437271$ systems along LOS
B, since \os\ aligned with \cf\ is discarded by the data (assuming same
\os/\cf\ ratio as in A). \os-bearing gas outflow would require detection of
\os\ aligned with \cf\ in B too, which is definitely not the case.

\subsection{Distribution of metals}

Correlations between galaxies and metal systems at $z=2-3$ have recently shown
that \os\ is always found at distances of up to a few hundred kpc from an
observed galaxy (Adelberger et al. 2005; Simcoe et al. 2006).  Our constraints
on characteristic absorber sizes and relative abundances from photoionization
models {\it and} on sizes and kinematic properties from observed transverse
differences all fit well into a scenario of galaxy feedback. The extended
\os-phase we detect here, though quite homogeneous on kpc scales, apparently
is dominated by metals that have been put in place only recently by Type II
SNe, in contrast to -- or on top of -- an early wide spread by Pop III stars.

Other observations (e.g., Adelberger et al. 2005) have indicated that gas {\it
is} able to escape galaxies and pollute the IGM up to large distances, and
this process may result in an inhomogeneous metal enrichment as suggested by
simulations (e.g., Theuns et al. 2002, Aguirre \& Schaye 2005). We may have
detected a new signature of this process in the disturbed \cf\ that is
associated with the \os\ systems. However, the dissociation between \cf\ and
\os\ observed here (both along and across the LOS) is somewhat uncomfortable
within that simple picture and may also indicate that the possible 
outflows required to transport the metals have different time scales for the
different species (and thus ionization stages) , with \os\ having been ``in
place'' well before \cf. On the other hand, we do not see the signature of
shock-heated gas; if \os\ systems like the one at $z=2.437271$ are
representative of such violent processes, then why does the gas appear
predominantly photoionized?  What fraction of the observed \os\ indeed comes
from galaxy feedback -- in contrast to arising in pre-enriched gas -- remains
to be investigated for example with similar observations of binary QSOs to
probe larger scales.

\section{Summary and  conclusions}
\label{conclusions}

We have searched for \os\ systems in spatially-resolved high-resolution
  spectra of two lensed QSOs. Applying a well-defined searching method
  independently to each spectrum we have detected \os\ in $10$ intervening
  \ho\ systems.  Their redshifts range between $z=2.010$ and $2.633$, and
  their \os\ column densities between $10^{12.9}$ and $10^{13.9}$ \icm. We
  have used the information provided by differences and similarities across
  the lines of sight in the line parameters of various species to extract
  physical parameters via photoionization models. { Given the inhomogeneous
  character of the data --different S/N ratios and degrees of blending--, we
  have paid special attention to the observational errors. We have also
  discussed the uncertainties of the photoionization models.}  In addition to
  this analysis, in the Appendix we present $3$ associated systems that also
  show absorption by \os. Our conclusions for the intervening systems are as
  follows:

   \begin{enumerate}

      \item Every \os\ system detected in one LOS shows \os\ also in the
      adjacent LOS. Since the observed number density of \os\ is in agreement
      with previous studies  and since \ho\ interlopers are believed not to
      show strong variations on kpc-scales, potential \os\ systems in only one
      LOS are likely not missing from our sample.

      \item Within uncertainties, \nos\ shows { little} structure across
      the LOS, which probe the gas on transverse scales of $0.2$--$1.4$ \hkpc,
      and the data are consistent with a null fractional change on these
      scales. { Unfortunately, the low S/N in the \lya\ forest of the B
      sightlines does not allow us to probe fractional changes lower than
      $\sim 40\%$ and the upper error bars are in some cases consistent with
      high values. On the other hand, the more certain 'Class I' systems 
      all show low fractional changes, thus supporting an \os\
      coherence length in excess of 1 kpc. The situation for \cf\ is fully
      different. Six out of the 8 systems in which we detect \cf\ show a
      fractional change in this species that is large (up to 90\%) and
      significant.}

      \item \cf\ and \os\ systematically show velocity offsets between line
      centroids while in the case where \ct\ is detected, it seems to follow
      \cf. There are also significant velocity offsets in \cf\ between LOS A
      and LOS B, but none of the \os\ systems show such offsets. Doppler
      widths indicate photoionized gas, and these also do not seem to vary
      between LOS.

      \item The observed properties of species in the 2 LOS
      support a scenario in which an important
      fraction of \cf\ may not reside in the same volume as \os. This is
      consistent with photoionization models that predict clouds
      with unrealistically large sizes if
      both ions are taken into account. A two-phase model in which \cf\ arises
      both in a low and a high-ionization regime, with characteristic sizes of
      a few tens and a few hundreds of kpc, respectively, successfully
      explains the various LOS differences.

      \item The observed \os\ puts constraints also on abundances, and
      photoionization requires for these clouds CNO abundances consistent with
      gas processed in galaxies recently before the absorption
      epoch. This implies that the absorbing gas is likely dominated by
      some kind of galactic feedback and/or it occurs not too far away from
      star-forming regions, although the observed \os\ is of a very different
      nature than in the Galaxy. In particular, the systems at $z=2.052$ and
      $z=2.437$ are  
      consistent with the lines of sight crossing a metal-rich galactic
      outflow. However, from its kinematics across the LOS, \os\ does not seem
      to be transported in the expelled gas.

   \end{enumerate}

\begin{acknowledgements}

We have benefitted from conversations with Jacqueline Bergeron, Todd Tripp and
Jason X. Prochaska.  The Haardt \& Madau (1996) spectrum had been kindly
provided to us by Francesco Haardt. SL would like to also thank the ESO
Scientific Visitor Program for supporting a pleasant stay at ESO Headquarters,
where part of this work was done.  SL was partly supported by the Chilean {\sl
Centro de Astrof\'\i sica} FONDAP No. 15010003, and by FONDECYT grant N$^{\rm
o}1060823$. We are also indebted to the referee, Joop Schaye, for insightful
criticisms.

\end{acknowledgements}

\appendix
\section{Associated systems}
\label{assoc}

The following three systems also have clear detections of \os\ but their
redshifts are within $v < 3\,000$ \kms\ of the systemic redshift of the QSO,
so we regard them as ``associated'' systems (e.g., Barlow \& Sargent
1997). For purposes of completeness, we briefly describe them here, but do not
consider them in our photoionization models. Figures A.1 to A.3 show observed
transitions of \ho, \cf, and \os\ plotted in velocity scale with respect to
arbitrary redshifts. It is important to note that the line-of-sight
separation, $S$, is computed in each case assuming that the redshift is purely
cosmological, i.e., $S$ is an upper limit if the absorption is produced in
outflowing material.

\paragraph{$z=2.300$, $S=100$ pc}

This system lies $v=1\,800$ \kms\ from the systemic redshift of \he.  We
fitted 5 velocity components to the \os\ systems along the A
line-of-sight. The $1037$ transition appears underfitted due to -- we believe
-- shallow absorption by \ho\ interlopers. Despite this caveat, a misaligment
between \os\ and \cf\ becomes clear in the 3 bluemost velocity components. 
{ Observations of a similar associated system have led Ganguly et al. (2006)
to postulate different gas phases with distinct degrees of ionization.  When
comparing with QSO B, \cf\ shows transverse structure while the \os\ does not,
which supports the scenario of different gas regions.}

This system is also observed in \sif, \sit, and \ct\ (not shown in the
Fig.) and is thus particularly well suited for studying physical conditions
of the absorbing gas. Assuming the QSO ionizes each of the individual clouds,
the size information provided by the two lines-of-sight can give
important constraints to the conditions and location of the absorbers within
the QSO host galaxy.

\paragraph{$z=2.315$, $S=25$ pc}

This system lies only $v=450$ \kms\ from the systemic redshift of \he.
All transitions appear identical between the two LOS. In
particular, the \os\ is weak and the doublet lines have the proper strength
ratio.

\paragraph{$z=2.777$, $S=70$ pc}

This system lies $v=2\,400$ \kms\ from the systemic redshift of \rx\ and shows
transitions of \os, \cf, \nf, and \ct\ (the latter two are not shown in
Fig.~\ref{fig_spec10}).  The system is composed of two ``clumps'' of \cf\
clouds around $v=0$ and $v=150$ \kms.  Only for the bluemost of them 
can a clear
identification of \os\ be made. This clump, however, shows no \ho, which
appears only at the redshift of the redmost components.  In the Fig. we plot
the two \ho\ (unsaturated) transitions used to constrain \nhi\ in A.  These
transitions are too noisy in B to perform a fit but comparison of low order
(saturated) Lyman lines suggests the \nhi\ in B should be similar as in A.

The fits attempted to the metal line profiles show clearly that the \os\
doublet lines in the blue components do not show the proper strength ratio,
and the same is true for \nf\ and \cf. This effect, seen along {\it both} LOS,
is only evident in the blue components; the two redmost \cf\ lines near $v=150$
\kms\ can perfectly be fitted.

Anomalous ratios in associated systems have been interpreted as produced by
absorbing clouds that do not fully cover the background continuum emission
region of the QSO (Barlow \& Sargent 1997). However, incomplete coverage would
be surprising in our case given the similar column densities in \os\ and \cf\
along LOS A and B, separated by much larger distances than the size of the
continuum source (the same applies if the excess flux is due to the unresolved
lensed images). The facts that the effect is observed along both LOS and that
the red clump does not show anomalous ratios in neither LOS rule out two other
possible explanations namely, emission by the lensing galaxy or a bad
continuum placement.

\begin{table}
\begin{minipage}[t]{18cm}
\caption{Associated \os\ systems}\label{table_assoc}      
\begin{tabular}{lcccc}
\hline\hline   
 Ion   & $z$      &$\log N$(A)\footnote{Total column densities.} &$\log N$(B)$^a$   &QSO\\ 
\hline
 {\ho} & 2.300 & 16.36(0.23)\footnote{From Lopez et al. (1999)}&16.53(0.23)$^a$&\he\\ 
 {\cf} &       & 14.01(0.02) & 14.32(0.02) &\\
 {\os} &       & 14.69(0.07)& 14.73(0.03) &\\
\hline
 {\ho} & 2.314 & 15.80(0.01)&15.33(0.04)&\he\\ 
 {\cf} &       & 12.77(0.01) & 12.82(0.01) &\\
 {\os} &       & 13.32(0.02)& 13.38(0.04) &\\
\hline
 {\ho} & 2.776 & 15.85(0.05)&...&\rx\\ 
 {\cf} &       & 14.10(0.03) &14.20(0.10)  &\\
 {\os} &       & 14.77(0.02)&14.75(0.03)  &\\
\hline
\end{tabular}
\end{minipage}
\end{table}


   \begin{figure}
   \centering
   \includegraphics[width=9cm,clip]{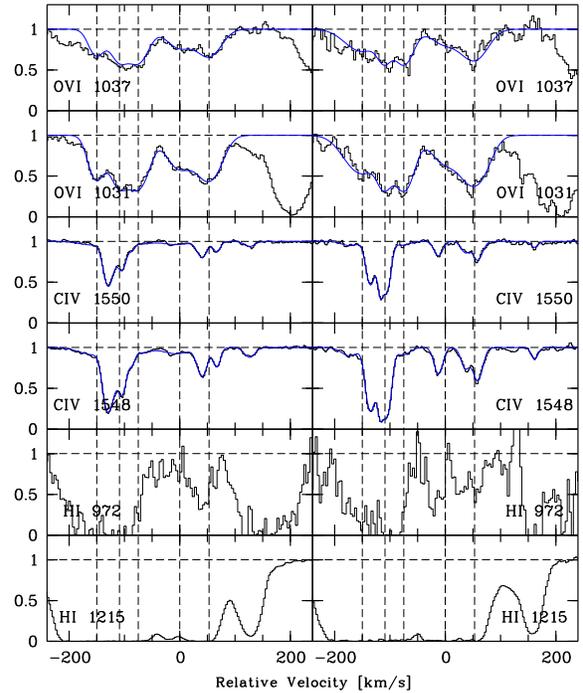}
   \caption{Associated system in \he\ at $z=2.300$.
Panels and symbols as for Fig.~\ref{fig_spec0}, but for   
   $z=2.298472$, $z=2.298920$, $z=2.299304$, $z=2.300127$, and
$z=2.300711$. 
}   
   \label{fig_spec8} 
    \end{figure}
   \begin{figure}
   \centering
   \includegraphics[width=9cm,clip]{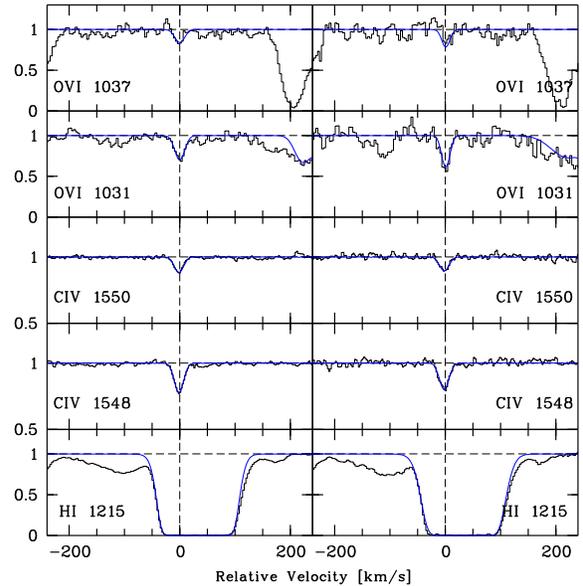}
   \caption{Associated system in \he\ at $z=2.314$.
Panels and symbols as for Fig.~\ref{fig_spec0}, but for   
   $z=2.314201$.}   
   \label{fig_spec9} 
    \end{figure}
   \begin{figure}
   \centering
   \includegraphics[width=9cm,clip]{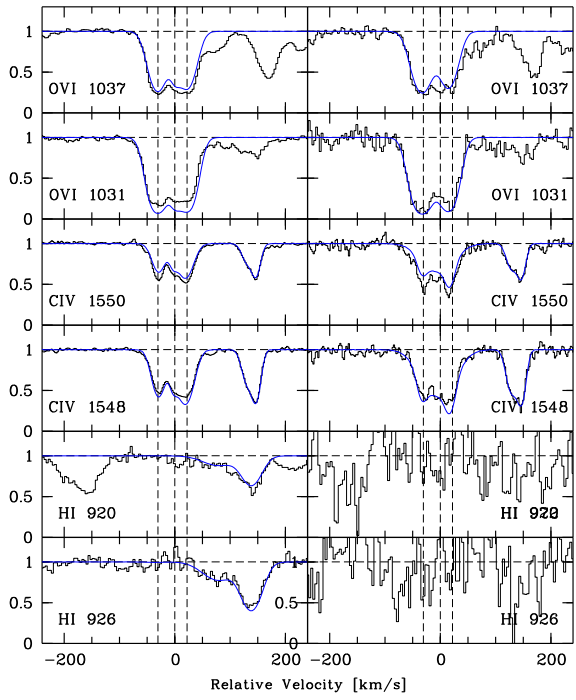}
   \caption{Associated system in \rx. Panels and symbols as for
   Fig.~\ref{fig_spec0}, but    for    
   $z=2.776014$, $z=2.776399$, and $z=2.776677$.}  
   \label{fig_spec10} 
    \end{figure}

\end{document}